\nonstopmode
\documentclass[twocolumn,showpacs,prb,amsmath,amssymb,floatfix,10pt,aps,eqsecnum,superscriptaddress]{revtex4-1}

\usepackage{amsmath,amssymb,color}
\usepackage{bm}
\usepackage{graphicx}
\usepackage{psfrag}
\usepackage{mhchem}
\usepackage{tabularx}
\usepackage{multirow}

\newcommand{\bd}{\bm}

\begin{document}

\title{Thermodynamics and renormalized quasi-particles in the vicinity of the dilute Bose gas quantum critical point in two dimensions}
 
\author{Jan Krieg}
\email{jkrieg@itp.uni-frankfurt.de}
\affiliation{Institut f\"{u}r Theoretische Physik, Universit\"{a}t Frankfurt,
Max-von-Laue Strasse 1, 60438 Frankfurt, Germany}

\author{Dominik Strassel}
\affiliation{Department of Physics and Research Center Optimas, University of Kaiserslautern, 67663 Kaiserslautern, Germany}

\author{Simon Streib}
\affiliation{Institut f\"{u}r Theoretische Physik, Universit\"{a}t Frankfurt,
Max-von-Laue Strasse 1, 60438 Frankfurt, Germany}
\affiliation{Kavli Institute of NanoScience, Delft University of Technology, Lorentzweg 1, 2628 CJ Delft, The Netherlands}

\author{Sebastian Eggert}
\affiliation{Department of Physics and Research Center Optimas, University of Kaiserslautern, 67663 Kaiserslautern, Germany}

\author{Peter Kopietz}
\affiliation{Institut f\"{u}r Theoretische Physik, Universit\"{a}t Frankfurt,
Max-von-Laue Strasse 1, 60438 Frankfurt, Germany}

\date{January 18th, 2017}

\begin{abstract}
We use the functional renormalization group (FRG) to derive analytical expressions for 
thermodynamic observables (density, pressure, entropy, and compressibility) as well as for single-particle properties 
(wavefunction renormalization and effective mass) of interacting bosons 
in two dimensions
as a function of temperature $T$ and chemical potential $\mu$.
We focus on the quantum disordered and the quantum critical regime close to the dilute Bose gas quantum critical point.
Our approach is based on a truncated vertex expansion of the hierarchy of FRG flow equations 
and the decoupling of the two-body contact interaction in the particle-particle channel using a suitable 
Hubbard-Stratonovich  transformation. 
Our analytic FRG results extend previous analytical renormalization group calculations for
thermodynamic observables at $\mu =0$
to finite values of $\mu$. 
To confirm the validity of our FRG approach, we have also performed quantum Monte Carlo simulations to obtain the magnetization, the susceptibility, and the correlation length of the two-dimensional spin-$1/2$ quantum $XY$ model with coupling $J$ in a regime where its quantum critical 
behavior is controlled by the dilute Bose gas quantum critical point. 
We find that our analytical results describe the Monte Carlo data for $\mu \leq 0$
rather accurately up to relatively high temperatures $T \lesssim 0.1 J$.

\end{abstract}


\maketitle

\section{Introduction}
\begin{figure}[h]
\includegraphics[width=\linewidth]{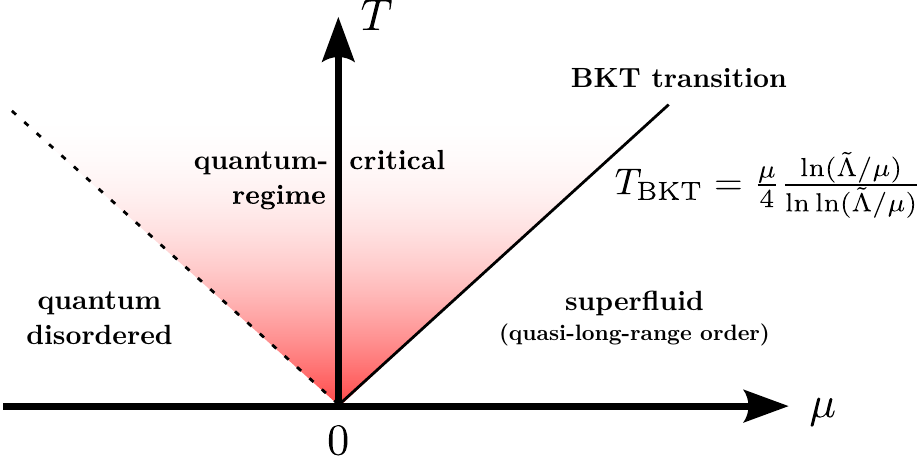}
\caption{
(Color online)
Schematic phase diagram for a two-dimensional Bose gas with repulsive contact interaction in the $T-\mu$ plane close to the QCP at $\mu = T = 0$. In the quantum disordered regime, $\mu < -T$, interaction effects are weak and the particle density is exponentially small, while for sufficiently large positive $\mu$ the system is in the superfluid phase with finite density even at zero temperature. This phase is separated from the normal phase by a BKT transition at the temperature $T_{\text{BKT}}$ as given in the figure, where $\tilde{\Lambda}$ is a non-universal energy scale.\cite{Pop72} In the quantum critical regime (red color) a quasi-particle description with free bosons is still valid, but physical quantities
exhibit logarithmic corrections.
}
\label{fig:phase_diagram}
\end{figure}

It is well known \cite{Sac11} that interacting bosons exhibit a quantum critical point (QCP) 
at vanishing chemical potential $\mu$ and temperature $T$ which separates 
a quantum disordered phase at $\mu < 0$ from a superfluid phase at $\mu > 0$
as sketched in Fig.~\ref{fig:phase_diagram}.
Of particular interest is the behavior of the system in two dimensions, where
the formation of a Bose-Einstein condensate for positive $\mu$ is prevented by the strong thermal fluctuations \cite{Mer66}. 
However, as has been shown by Popov,\cite{Pop72} the system nevertheless exhibits a phase transition of the Berezinsky-Kosterlitz-Thouless (BKT) type\cite{Ber71,Ber72,Kos73,Kos74} into a superfluid phase 
with quasi-long-range order and algebraically decaying  correlations. 
This is closely related to the fact that the upper critical dimension at the QCP is $D_c = 2$, 
so that the contact interaction is marginal in the renormalization group
sense:\cite{Fis89} 
while directly at the QCP the interaction is renormalized to vanishing strength, 
the renormalization group flow approaches zero only logarithmically. 
As a result, at any realistic temperature the collective dynamics of the system is strongly coupled, 
even though the effective pairwise interaction may be weak.\cite{Sac06} 
Hence perturbation theory is not applicable at finite $T$ and one has to resort to non-perturbative methods to analyze the superfluid as well as the quantum critical regime. 
The renewed theoretical interest in this model is motivated by a multitude 
of cold atoms experiments \cite{Had06,Kru07,Cla09,Tun10,Hun11,Yef11,Zha12,Ha13,Des14,Fle15} which have explored the phase diagram as well as the BKT transition in two-dimensional Bose gases.

In early theoretical works \cite{Pop72,Fis88,Sac94}
the properties of the superfluid phase and the position of the BKT transition 
were investigated in the extremely dilute limit where the effective dimensionless interaction $g$ is renormalized to very small values [see Eq.~(\ref{eq:g_def}) below]. However, for quantitative calculations this limit may not be realistically reached as was demonstrated by Prokof'ev {\it{et al.}} \cite{Pro01,Pro02} since, e.g., the critical density $n_c$ at the BKT transition exhibits a logarithmic dependence of the form
\begin{equation}
n_c = \frac{m T}{2 \pi} \ln \frac{C}{g},
\label{eq:critical_density_intro}
\end{equation}
while the effective interaction $g$ in turn only depends logarithmically on the density (due to $D = D_c$). 
Here $m$ is the mass of the bosons and we use units where $\hbar = k_B = 1$. The constant $C$ appearing in the 
logarithm in Eq.~(\ref{eq:critical_density_intro}) has been computed numerically using the classical $|\phi|^4$ model 
as $C \approx 121$,\cite{Pro01} hence the necessary limit $\ln (1/g) \gg \ln C$ where $C$ can be neglected is not accessible 
with available experimental techniques. Nonetheless it is possible to reach relatively small $g$ of the order of $0.01$ 
experimentally, e.g., by tuning the bare interaction of harmonically trapped atoms to small values through a magnetic Feshbach resonance.\cite{Hun11,Ha13,Fle15} Further theoretical works also increasingly employed numerical methods to complement the asymptotic analysis.\cite{Ber02,Pil05,Sac06,Ran12,Cec13,St15}

Here we are mainly interested in the universal scaling inside the quantum critical regime where $T \gtrsim |\mu|$. For the special case $\mu = 0$ this has already been investigated analytically by Sachdev {\it{et al.}} \cite{Sac94}. Recently Ran\c con and Dupuis \cite{Ran12} expanded on this by using a
functional renormalization group (FRG) approach based on a truncated gradient expansion.
By solving the truncated FRG flow equations  numerically they calculated the 
scaling of state functions close to the QCP for vanishing and finite chemical potential; 
they also reconsidered the analytical behavior for $\mu = 0$ and corrected the result for the density in Ref.~[\onlinecite{Sac94}].
In this work we shall re-examine the universal scaling within the FRG, using an alternative
truncation strategy of the formally exact hierarchy of FRG flow equations 
based on the vertex expansion.\cite{Kop10}
This enables us to extend
previous analytical results for the universal scaling in the quantum critical regime.
We also compare  our FRG results to quantum Monte Carlo (QMC) simulations. 

The rest of this work is organized as follows: In Sec.~\ref{sec:flow_equations} we 
introduce a Hubbard-Stratonovich transformation to
decouple the contact interaction in the particle-particle channel and 
derive FRG flow equations for the self-energy and the particle-particle susceptibility. 
We then solve these flow equations approximately in Sec.~\ref{sec:thermodynamics} and obtain 
analytical expressions for the pressure, the density, the entropy, the compressibility, and the correlation length which are valid close to the 
QCP; we also compare these results to experimental as well as to numerical data. 
In
Sec.~\ref{sec:quantum_monte_carlo} we discuss the spin-1/2 quantum $XY$ model in two dimensions which can be 
mapped to the Bose-Hubbard model with infinite onsite interaction, hence showing the same universal scaling towards the QCP as the dilute Bose gas. We present results from QMC simulations of this spin system and compare them to our analytical FRG results. Finally in Sec.~\ref{sec:quasiparticle_properties} we use the FRG formalism to study the scaling of the effective mass and the wavefunction renormalization; we also compare the FRG result for the wavefunction renormalization
with the result from the self-consistent $T$-matrix approximation, which is applicable for $\mu < 0$ 
and at high temperatures also for $\mu = 0$. Further technical details can be found in two appendices:
in appendix A we present formally exact FRG flow equations for the irreducible vertex functions of 
interacting bosons, while in appendix B we give some mathematical details on the analytical
solution of our truncated FRG flow equations.

\section{FRG flow equations for dilute bosons}
\label{sec:flow_equations}

We consider a system of interacting bosons with
contact two-body interaction and second quantized Hamiltonian 
  \begin{eqnarray}
 {\mathcal{H}} & = & \sum_{\bd{k}} \epsilon_{\bd{k}} {\hat{a}}^{\dagger}_{\bd{k}}
  {\hat{a}}_{\bd{k}}
 + \frac{f_0}{4V} \sum_{ \bd{p} \bd{k} \bd{k}^{\prime}}
  {\hat{a}}^{\dagger}_{ \bd{p}  - \bd{k}    } 
 {\hat{a}}^{\dagger}_{\bd{k}} {\hat{a}}_{\bd{k}^{\prime}}
 {\hat{a}}_{ \bd{p} - \bd{k}^{\prime} },
 \label{eq:hamiltonian}
 \end{eqnarray}
where $\hat{a}_{\bd{k}}$ annihilates a boson with 
momentum $\bd{k}$, energy
$\epsilon_{\bd{k}} = \bd{k}^2 /(2m)$, and mass $m$.
The volume of the system is denoted by $V$ and the normalization of the contact
two-body interaction with strength $f_0$ has been chosen to simplify
the combinatorial factors in the FRG flow equations
given in appendix~\ref{sec:exact_frg_flow_equations}.
We have shifted the momentum labels 
in Eq.~(\ref{eq:hamiltonian})  such
that $\bd{p}$ can be identified with the conserved total momentum of a pair of
incoming or outgoing bosons. This labeling is natural in the dilute limit
where the particle-particle channel is the dominant scattering process.

\subsection{Hubbard-Stratonovich transformation in the
particle-particle channel}

At finite chemical potential $\mu$ and temperature $T  = 1 / \beta$,
the Euclidean action associated with the Hamiltonian (\ref{eq:hamiltonian})
can be written as
 \begin{equation}
 S [\bar{a} , a ]  =  - \int_K G_0^{-1} ( K ) \bar{a}_K a_K + \frac{f_0}{4} \int_P  \bar{A}_P A_P,
 \label{eq:Sdef}
 \end{equation}
where the free boson propagator is
 \begin{equation}
 G_0 ( K ) = \frac{1}{ i \omega - \epsilon_{\bd{k}} + \mu },
 \end{equation}
and we have introduced the composite boson fields
 \begin{equation}
 A_P = \int_K a_{K} a_{  P-K }, \quad \bar{A}_P = \int_K \bar{a}_{ P - K} \bar{a}_{K }.
 \label{eq:composite2}
 \end{equation}
Here $K = ( \bd{k} , i \omega ) $ and $P = ( \bd{p} , i \bar{\omega} )$ are collective
labels for momenta and bosonic Matsubara frequencies, the integration symbols
are defined by $\int_K = \frac{1}{\beta V } \sum_{\bd{k} , \omega }$, 
and $a_{K}$ is a complex field associated with the eigenvalues of $\hat{a}_{\bd{k}}$.
Introducing another complex boson field $\psi_P$ to decouple
the interaction by means of a Hubbard-Stratonovich (HS) transformation
in the particle-particle channel we obtain
 \begin{align}
 \label{eq:action_hs}
 S [\bar{a} , a , \bar{\psi} , \psi ]  &=  - \int_K G_0^{-1} ( K ) \bar{a}_K a_K 
  + \int_P f_0^{-1} \bar{\psi}_P \psi_P 
 \nonumber 
 \\
 &+
\frac{i}{2!} \int_P [    \bar{A}_P  \psi_P  +    \bar{\psi}_P    A_P     ].
 \end{align}
We have normalized the 
$\psi$-field to simplify the
combinatorial factors in the exact FRG flow equations given in appendix~\ref{sec:exact_frg_flow_equations}.
Below we shall refer to the original boson fields $a$ and $\bar{a}$ as elementary bosons
and to the boson fields $\psi$ and $\bar{\psi}$ as HS bosons.

\subsection{Truncated FRG flow equations}
To set up the FRG, we introduce a sharp cutoff in momentum space for the elementary boson so that the regularized non-interacting propagator is given by
 \begin{equation}
 G_{0, \Lambda}( K )  = \frac{ \Theta ( | \bd{k} | - \Lambda ) }{ i \omega - \xi_{\bd{k}} },
 \end{equation}
and the corresponding single-scale propagator is
 \begin{equation}
 \dot{G}_{\Lambda} ( K ) = - \frac{\delta ( | \bd{k} | - \Lambda )}{ i \omega - \xi_{\bd{k}}
 - \Sigma_{\Lambda} ( K ) },
 \label{eq:Gdot}
 \end{equation}
where $\Sigma_\Lambda (K)$ is the cutoff dependent self-energy and we have defined $\xi_{\bd{k}} = \epsilon_{\bd{k}} - \mu$. For our purpose, it is sufficient to use 
the following ansatz for the generating functional of the
irreducible vertices,
\begin{align}
 \Gamma_{\Lambda} [\bar{a} , a , \bar{\psi} , \psi ] &=
 \int_K \Sigma_{\Lambda} ( K ) \bar{a}_K a_K + \int_P \Pi_{\Lambda} ( P ) \bar{\psi}_P \psi_P
 \nonumber
 \\
 & \hspace{-10mm} + \frac{1}{2!} \int_K \int_P  \Bigl[
 \Gamma_{\Lambda}^{ \bar{a} \bar{a} \psi} (P-K, K; P ) \bar{a}_{ P-K } \bar{a}_K \psi_P
 \nonumber
 \\
 & \hspace{5mm} +  \Gamma_{\Lambda}^{ {a} {a} \bar{\psi}} (P-K,  K; P ) {a}_{ P-K } {a}_K \bar{\psi}_P
 \Bigr],
 \label{eq:ansatz_generating_functional}
\end{align}
where the energy-momentum labels of the three-legged vertices $ \Gamma_{\Lambda}^{ \bar{a} \bar{a} \psi} (P-K,   K; P )$ 
and $\Gamma_{\Lambda}^{ {a} {a} \bar{\psi}} (P-K,  K; P )$ correspond to the field types 
appearing in the superscripts.
This ansatz is justified since all four-point and higher order vertices are irrelevant in the RG sense, except for $\Gamma_\Lambda^{\bar{a}\bar{a}aa}$ which is marginal. This vertex, however, initially vanishes due to the HS transformation and is only dynamically generated during the flow by either particle-hole processes or by vertices which are irrelevant in the RG sense (see Fig.~\ref{fig:diagrams_four}). As our calculations are concerned with the dilute limit, particle-hole processes are suppressed and the relevant physics is captured by particle-particle processes via the three-point vertices $ \Gamma_{\Lambda}^{ \bar{a} \bar{a} \psi}$ and $\Gamma_{\Lambda}^{ {a} {a} \bar{\psi}}$.
The exact FRG flow equations for the self-energy $\Sigma_\Lambda (K)$ and the particle-particle 
susceptibility $\Pi_\Lambda (P)$ as well as for all three- and four-legged vertices of our model 
are given in appendix~\ref{sec:exact_frg_flow_equations}. 
From the flow equations (\ref{eq:gamma3a}) and (\ref{eq:gamma3b}) for the three-legged vertices we see that within our ansatz (\ref{eq:ansatz_generating_functional}) for the generating functional these vertices do not flow, so that we can replace them by their initial value
 \begin{equation}
  \Gamma_{\Lambda}^{ \bar{a} \bar{a} \psi} (P-K,  K; P ) = 
\Gamma_{\Lambda}^{ {a} {a} \bar{\psi}} (P-K,  K; P ) = i.
 \end{equation}
The exact flow equation for the self-energy given in Eq.~(\ref{eq:flowself}) then simplifies to
\begin{equation}
 \partial_{\Lambda} \Sigma_{\Lambda} ( K ) =
 -   \int_P    {F}_{\Lambda} ( P ) \dot{G}_{\Lambda} ( P - K ),
 \label{eq:flowselfdilute}
 \end{equation}
while the flow equation for the particle-particle susceptibility given in Eq.~(\ref{eq:flowpol}) becomes
 \begin{equation}
 \partial_{\Lambda} \Pi_{\Lambda} ( P ) =
  \int_K  \dot{G}_{\Lambda} ( K ) G_{\Lambda} ( P - K ).
 \label{eq:flowpoldilute}
 \end{equation}
Here we have introduced the flowing propagator
\begin{equation}
F_\Lambda (P) = \frac{f_0}{1 + f_0 \Pi_\Lambda (P)}
\end{equation}
of the HS boson. We expect these equations to be accurate in the vicinity
of the dilute Bose gas QCP where
particle-hole scattering processes can be neglected.

Note that if we replace the single-scale propagators in Eqs.~(\ref{eq:flowselfdilute}) and (\ref{eq:flowpoldilute}) by total derivatives with respect to $\Lambda$ and ignore the $\Lambda$ dependence of the particle-particle susceptibility $\Pi_\Lambda$, we can integrate both sides of these equations over $\Lambda$ to obtain
\begin{align}
\Sigma (K) &= - \int_P \frac{f_0}{1 + f_0 \Pi (P)} G (P-K),
\label{eq:self_consistent_t_matrix_sigma}
\\*
\Pi (P) &= \frac{1}{2} \int_K G (K) G (P-K).
\label{eq:self_consistent_t_matrix_pi}
\end{align}
These coupled integral equations are usually called the self-consistent $T$-matrix approximation.
For an early application of this method to the dilute Bose gas in two dimensions see Ref.~[\onlinecite{Sto93}]. 
Some of us \cite{Str15} have recently used
this  approximation to study an effective hard-core boson model 
describing the magnetic properties of the antiferromagnetic material $\text{Cs}_2\text{CuCl}_4$ 
(see also Ref.~[\onlinecite{Fau14}] for recently discovered subtleties in this method when applied to hard-core bosons). In Sec.~\ref{sec:wavefunction_renormalization} we shall compare our FRG results for the wavefunction renormalization at $\mu = 0$ to the results obtained from the numerical solution of the integral equations (\ref{eq:self_consistent_t_matrix_sigma})
and (\ref{eq:self_consistent_t_matrix_pi}).

\section{Thermodynamics close to the QCP}
\label{sec:thermodynamics}

\subsection{RG flow at the quantum critical point}
\label{sec:exact_at_qcp}
To begin with, let us briefly recall the renormalization group (RG) flow
of the system directly at the QCP in $D$ dimensions. 
Since in this case the equilibrium state of the system corresponds to the vacuum, 
the elementary propagator $G_\Lambda (K)$ is identical to the free propagator $G_{0,\Lambda} (K)$, i.e., $\Sigma_\Lambda (K) = 0$. According to Eqs.~(\ref{eq:flowselfdilute}) and (\ref{eq:flowpoldilute}) the flow of the particle-particle susceptibility for vanishing momentum and frequency then simplifies to
\begin{equation}
\partial_\Lambda \Pi_\Lambda (0) = - \frac{K_D}{2} m \Lambda^{D-3},
\end{equation}
where $K_D$ is the surface area of the $D$-dimensional unit sphere divided by $(2 \pi)^D$. Defining the dimensionless rescaled interaction
\begin{equation}
u_\Lambda = \frac{K_D}{2} m \Lambda^{D-2} F_\Lambda (0)
 \label{eq:uLambdadef}
\end{equation}
and switching to the logarithmic scale parameter $l = \ln (\Lambda_0 / \Lambda)$, where $\Lambda_0$ is the ultraviolet cutoff of our theory, we arrive at the well-known exact flow equation\cite{Fis88,Sac11}
\begin{equation}
\partial_l u_l = (2 - D) u_l - u_l^2,
\end{equation}
which identifies $D_c = 2$ as the upper critical dimension above which mean field theory is applicable. In the following we will always work at $D = D_c$, resulting in logarithmic corrections to the scaling of various observables.

\subsection{Explicit solution of the FRG equations close to the quantum critical point in two dimensions}
\label{sec:frg_qcp}
The above system of FRG flow equations (\ref{eq:flowselfdilute}) and (\ref{eq:flowpoldilute}) for the self-energy and the particle-particle susceptibility can be solved approximately by neglecting their momentum and frequency dependence which is justified close to the QCP. In the following we will again use the scaling parameter $l = \ln (\Lambda_0 / \Lambda)$ and the dimensionless interaction
\begin{equation}
u_l = \frac{m}{4 \pi} F_l (0),
\end{equation}
which is a special case of Eq.~(\ref{eq:uLambdadef}) for $D=2$.
We also define the effective inverse temperature
\begin{equation}
\beta_l = \frac{\Lambda^2}{2 m T} = \frac{\Lambda_0^2}{2 m T} e^{-2l},
\end{equation}
the dimensionless particle-particle susceptibility
\begin{equation}
\tilde{\Pi}_l = \frac{4 \pi}{m} \Pi_l (0),
\end{equation}
and the effective negative chemical potential over temperature
\begin{equation}
r_l = - \frac{[\mu - \Sigma_l (0)]}{T} = \alpha + \frac{\Sigma_l (0)}{T},
\end{equation}
where we have introduced the abbreviation
\begin{equation}
\alpha = - \frac{\mu}{T}
\label{eq:alpha}
\end{equation}
to simplify the forthcoming equations. 
With this notation
the flow equations (\ref{eq:flowselfdilute}) and (\ref{eq:flowpoldilute}) 
can be written as
\begin{align}
\partial_l r_l &= \frac{u_0}{1 + u_0 \tilde{\Pi}_l} \frac{4 \beta_l}{e^{\beta_l + r_l} - 1},
\label{eq:flow_equation_r_self_energy}
\\*
\partial_l \tilde{\Pi}_l &= \frac{\beta_l}{\beta_l + r_l} \left[1 + \frac{2}{e^{\beta_l + r_l} - 1} \right].
\label{eq:flow_equation_pi_self_energy}
\end{align}
It turns out that we can analytically solve this system of 
differential equations approximately for $r = \lim_{l \rightarrow \infty} r_l$ 
provided we 
consider the regime close to the QCP and the following dimensionless coupling is sufficiently small,
\begin{equation}
g = \frac{2}{\frac{1}{u_0} + \frac{1}{2} \ln \left[ \frac{\Lambda_0^2}{2 m (T - \mu)} \right] },
\label{eq:g_def}
\end{equation}
where the logarithmic term results from the flow of $\tilde{\Pi}_l$ [cf. Eq.~(\ref{eq:boundary_condition_new})].
Technical details on the analytical solution of
Eqs.~(\ref{eq:flow_equation_r_self_energy}) and (\ref{eq:flow_equation_pi_self_energy}) 
are given in appendix~\ref{sec:detailed_calcs_self_energy}.
There we show that an approximate analytical solution is
possible if
either $g \ll \alpha$ or $W(1/g) \gg 1$, where $W(x)$ denotes the Lambert W function \cite{Cor96} which for large arguments can be expanded as
\begin{equation}
W(x) = \ln x - \ln \ln x + o(1).
\end{equation}
The condition $W(1/g) \gg 1$ is thus fulfilled either extremely close to the QCP or for small 
bare interaction $u_0$. 
The regime $u_0 \ll 1$ is of significant practical importance as it allows to probe the asymptotic scaling behavior experimentally by tuning the interaction to small values, since reaching double exponentially low temperatures is not feasible. Actually for small $u_0$ the weak logarithmic dependence on $T$ results in an approximate temperature independence of $g$.

As shown in appendix~\ref{sec:detailed_calcs_self_energy}, for $\mu \leq 0$ the limit $r = \lim_{l \rightarrow \infty} r_l$ can be written as
\begin{equation}
r = g W \left[ \frac{1}{g} \exp \left( \frac{e^\alpha - 1}{g} + \alpha \right) \right] - e^{\alpha} + 1 + \alpha,
\label{eq:sigma_result_general}
\end{equation}
which in the special case $\mu = 0$ simplifies to
\begin{equation}
r = g W (1/g).
\label{eq:sigma_mu_0}
\end{equation}
On the other hand, for finite $\alpha > 0$ and $g \ll 1$ we can expand Eq.~(\ref{eq:sigma_result_general}) as
\begin{align}
r &= \alpha + g \ln \left( \frac{1}{1 - e^{-\alpha}} \right) - \frac{g^2}{e^\alpha - 1} \ln \left( \frac{1}{1 - e^{-\alpha}} \right) + \mathcal{O} (g^3),
\label{eq:expansion_r_finite_alpha}
\end{align}
which in the quantum disordered regime where $\alpha \gg 1$ results in
\begin{equation}
r = \alpha + g e^{-\alpha} - \left( g e^{-\alpha} \right)^2 + \mathcal{O} (g^3),
\end{equation}
so that the self-energy is exponentially suppressed.

In appendix~\ref{sec:detailed_calcs_self_energy} we also derive
a more general expression,
\begin{align}
r &= \frac{2 g}{2 - 3 g} W \left[ \frac{2 - 3 g}{2 g} \exp\left( \frac{2 - 3 g}{2 g} \left[ 1 - e^{-\alpha} \left( 1 - \frac{g}{2} \right) \right] \right) \right]
\nonumber
\\*
& \hspace{7mm} - 1 + e^{-\alpha} \left( 1 - \frac{g}{2} \right) + \alpha,
\label{eq:sigma_result_general_posMu}
\end{align}
which is valid for arbitrary $\alpha = - \mu / T$ as long as we stay in the normal phase close to the QCP and do not come too close to the BKT phase transition.
The reason for the latter constraint is that vortices become increasingly important close to the superfluid transition which eventually leads to a breakdown of the quasi-particle approximation $\Sigma(K) \approx \Sigma(0)$ (cf. Fig.~\ref{fig:z_factor_absolute} at $\mu > 0$). Nevertheless, extrapolating (\ref{eq:sigma_result_general_posMu}) into the classical critical regime we can give an estimate for the critical chemical potential $\mu_c$ at the BKT transition by demanding that $r = 0$, which in the asymptotic limit yields
\begin{equation}
\frac{\mu_c}{T} = g \ln \frac{2}{g}.
\label{eq:phase_transition_g}
\end{equation}
Given that our theory is not justified in the classical critical regime, this agrees well with the weak-coupling result \cite{Pop83,Fis88}
\begin{equation}
\frac{\mu_c}{T} = g \ln \frac{C_\mu}{g},
\end{equation}
where $C_\mu$ has been obtained numerically as $C_\mu \approx 4.2$ in Ref.~[\onlinecite{Pro01}] and as $C_\mu \approx 3.0$ in Ref.~[\onlinecite{Ran12}].

\subsection{Thermodynamic state functions}

Within our approximation scheme the density at scale $\Lambda$ is given by
\begin{equation}
n_{\Lambda} = \int \frac{ d^2 k}{(2 \pi )^2} \frac{\Theta ( | \bd{k} | - \Lambda) }{
e^{ [ \epsilon_{\bd{k}} + \Sigma_{\Lambda} (0) - \mu ] / T } -1 },
\end{equation}
which corresponds to the particle density of all particles with momentum $k > \Lambda$.
The integration can be carried out exactly and we obtain for the 
physical density
\begin{equation}
n = \lim_{\Lambda \to 0} n_\Lambda = - \frac{m T}{2 \pi} \ln \left[ 1 - e^{ -  r}  \right].
\label{eq:density_gen_result_2D}
\end{equation}
Analogously the off-diagonal elements of the density matrix in our approximation are
\begin{equation}
G_\Lambda (\bd{x},\bd{x}') = \int \frac{d^2 k}{(2 \pi)^2} \frac{e^{i \bd{k} \cdot (\bd{x} - \bd{x}')} \Theta ( | \bd{k} | - \Lambda) }{
e^{ [ \epsilon_{\bd{k}} + \Sigma_{\Lambda} (0) - \mu ] / T } -1 }.
\end{equation}
In the limit of large distances $|\bd{x} - \bd{x}'| \to \infty$  and for $\Lambda \to 0$ we can evaluate the momentum integration analytically,\cite{Kop10}
\begin{equation}
G (\bd{x},\bd{x}') = \lim_{\Lambda \to 0} G_\Lambda (\bd{x},\bd{x}') \sim \frac{e^{- |\bd{x} - \bd{x}'| / \xi}}{\sqrt{|\bd{x} - \bd{x}'| / \xi}},
\end{equation}
where we have introduced the correlation length
\begin{equation}
\xi = 1 / \sqrt{2 m T r}.
\label{eq:correlationlength}
\end{equation}
We now define the reduced pressure $\tilde{p}$, the phase-space density $\tilde{n}$, the entropy per particle $\tilde{s}$, the dimensionless compressibility $\tilde{\kappa}$, and the dimensionless correlation length $\tilde{\xi}$,
\begin{subequations}
\begin{align}
\tilde{p} &= \frac{\lambda_{\text{th}}^2}{T} p = - \frac{\lambda_{\text{th}}^2}{T} \frac{\Omega}{V},
\\
\tilde{n} &= \lambda_{\text{th}}^2 n,
\\
\tilde{s} &= \frac{1}{n} \frac{S}{V},
\\
\tilde{\kappa} &= \frac{2 \pi}{m} \kappa = - \left( \frac{\partial \tilde{n}}{\partial \alpha} \right)_T,
\label{eq:kappa}
\\
\tilde{\xi} &= \sqrt{2 m T} \xi,
\end{align}
\end{subequations}
where $p$ is the pressure, $\Omega$ is the grand canonical potential, $S$ is the entropy, 
$\kappa = \left(\partial n / \partial \mu \right)_T$ is the compressibility, and
the thermal de Broglie wavelength is given by
\begin{equation}
\lambda_{\text{th}} = \sqrt{\frac{2 \pi}{m T}}.
\end{equation}
Since we have approximated the self-energy by its zero momentum and frequency limit, we can try to incorporate the interaction solely as a shift in the chemical potential, neglecting the renormalized interaction $u$. Thus we compute the state functions for a non-interacting Bose gas and fix the chemical potential such that the particle density $n_{\text{free}}$ coincides with Eq.~(\ref{eq:density_gen_result_2D}), i.e., $\mu_{\text{free}} = \mu - \Sigma (0)$. Accordingly we find
\begin{subequations}
\begin{align}
\label{eq:result_pressure_free}
\tilde{p} &= \text{Li}_2 (e^{-r}) \underset{\mu = 0}{\sim} \frac{\pi^2}{6} - g W^2 \left(1/g\right),
\\*
\label{eq:result_density_free}
\tilde{n} &= \ln \left[ \frac{1}{1 - e^{-r}} \right] \underset{\mu = 0}{\sim} W(1/g),
\\*
\label{eq:result_entropy_free}
\tilde{s} &= \frac{2 \tilde{p}}{\tilde{n}} + r \underset{\mu = 0}{\sim} \frac{\pi^2}{3 W(1/g)},
\\*
\label{eq:result_correlation_length}
\tilde{\xi} &= 1 / \sqrt{r} \underset{\mu = 0}{\sim} 1 / \sqrt{g W(1/g)},
\end{align}
\end{subequations}
where $\text{Li}_2 (x)$ is the dilogarithm. Calculating the dimensionless compressibility from Eq.~(\ref{eq:result_density_free}) yields
\begin{equation}
\tilde{\kappa} = \frac{ \left( \frac{\partial r }{ \partial \alpha}\right)_T }{e^r - 1} \underset{\mu = 0}{\sim} \frac{1}{g W(1/g)}.
\label{eq:kappa-leading}
\end{equation}
We note that these observables do not have a well-defined limit at $\mu = T = 0$ due to the non-analyticity of the grand canonical potential at the QCP, which separates the zero density ground state at $\mu < 0$ from the finite density superfluid ground state at $\mu > 0$.
For the special case of vanishing chemical potential 
the above relations have 
already been obtained  by Ran\c con and Dupuis \cite{Ran12} whose results
agree with our expressions for $\mu=0$. 

Note that Eq.~(\ref{eq:result_density_free})  
corrects the result for the density at $\mu = 0$ given by Sachdev {\it{et al.}},\cite{Sac94}
\begin{equation}
n = \frac{m T}{2 \pi} \ln^{-4} \left( \frac{\Lambda_0^2}{2 m T} \right), \quad (\text{Ref.~[\onlinecite{Sac94}]})
\end{equation}
while we find from Eqs.~(\ref{eq:result_density_free}) and (\ref{eq:g_def})
\begin{align}
n &= \frac{m T}{2 \pi} W \left[ \frac{1}{2 u_0} + \frac{1}{4} \ln \left( \frac{\Lambda_0^2}{2 m T} \right) \right]
\nonumber
\\
&\underset{T \to 0}{\sim} \frac{m T}{2 \pi} \ln \left[ \frac{1}{4} \ln \left( \frac{\Lambda_0^2}{2 m T} \right) \right].
\label{simplelog}
\end{align}

We can improve on the calculation of $\tilde{p}$ and $\tilde{s}$ by directly calculating the grand canonical potential within the FRG formalism, solving the flow equation
\begin{equation}
\frac{\partial_\Lambda \Omega_\Lambda}{V} = - \int_K \frac{\dot{G}_{0,\Lambda} (K) \Sigma_\Lambda (K)}{1 - G_{0,\Lambda} (K) \Sigma_\Lambda (K)}.
\end{equation}
Allowing for first order corrections in the self-energy,
\begin{equation}
\Sigma_\Lambda (K) \approx \Sigma_\Lambda (0) - (1 - Y_\Lambda^{-1}) \epsilon_{\bd{k}} + (1 - Z_\Lambda^{-1}) i \omega,
\label{eq:self_energy_expansion}
\end{equation}
the flow equation for the reduced pressure $\tilde{p} = - \frac{\lambda_{\text{th}}^2}{T} \frac{\Omega}{V}$ reads within our cutoff scheme (see appendix~\ref{sec:detailed_free_energy})
\begin{align}
\partial_l \tilde{p}_l = &- 2 \beta_l \Biggl[ \ln \left( \frac{e^{Z_l (Y_l^{-1} \beta_l + r_l)} - 1}{e^{\beta_l + \alpha} - 1} \right)
\nonumber
\\*
&  \hspace{10mm} - Z_l (Y_l^{-1} \beta_l + r_l) + \beta_l + \alpha \Biggr].
\label{eq:flowEquationReducedPressure}
\end{align}
Approximating $Z_l = Y_l = 1$ and $r_l = r$ results in $\tilde{p} = \text{Li}_2 (e^{-r})$ which agrees with the reduced pressure in Eq.~(\ref{eq:result_pressure_free}). In contrast, if we solve the flow equation \eqref{eq:flowEquationReducedPressure} for $\tilde{p_l}$ at $\mu = 0$ with the flowing $r_l$ we find (see appendix~\ref{sec:detailed_free_energy_mu_0})
\begin{equation}
\tilde{p} \sim \frac{\pi^2}{6} - \frac{g}{2} W^2(1/g).
\end{equation}
Here the leading correction is only half as large as our earlier result in Eq.~(\ref{eq:result_pressure_free}). This implies that it is necessary to solve the flow equation for the grand canonical potential to obtain the correct leading order scaling behavior for the reduced pressure. The entropy per particle, which at $\mu = 0$ can be derived from the reduced pressure as
\begin{equation}
\tilde{s} = \frac{1}{\tilde{n}} \left( 2 \tilde{p} + \frac{g^2}{4} \frac{\partial \tilde{p}}{\partial g} \right),
\label{eq:entropy_equation_from_flow}
\end{equation}
is only affected to subleading order. While the phase-space density can in principle also be calculated from $\tilde{p}$ as $\tilde{n} = \partial \tilde{p} / \partial (\mu/T)_T$, it is preferable to use Eq.~(\ref{eq:density_gen_result_2D}) as it avoids the additional approximations in the computation of the reduced pressure.

\begin{figure}
\includegraphics[width=0.975\linewidth]{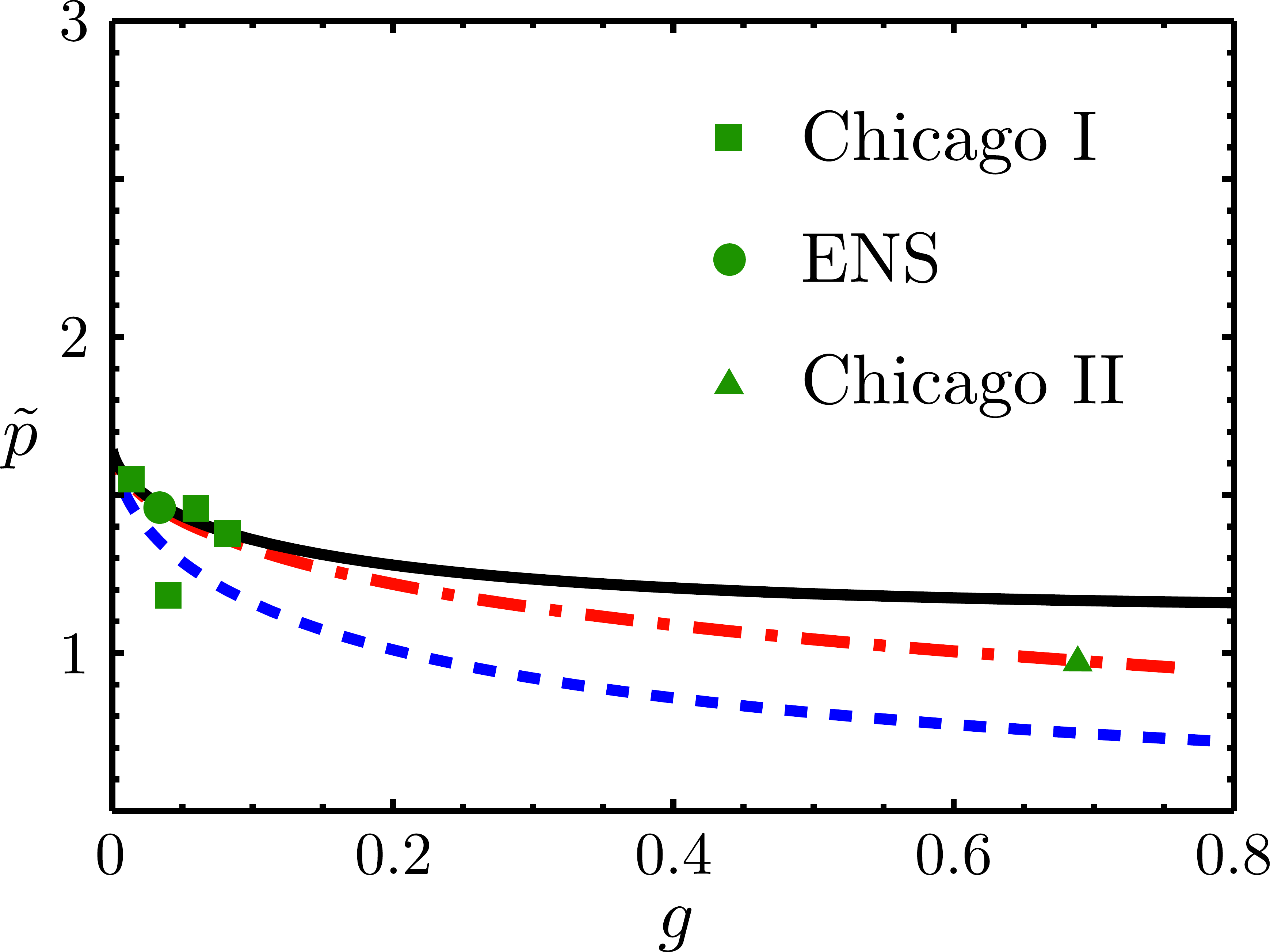}

%\vspace{2.5mm}

\includegraphics[width=0.975\linewidth]{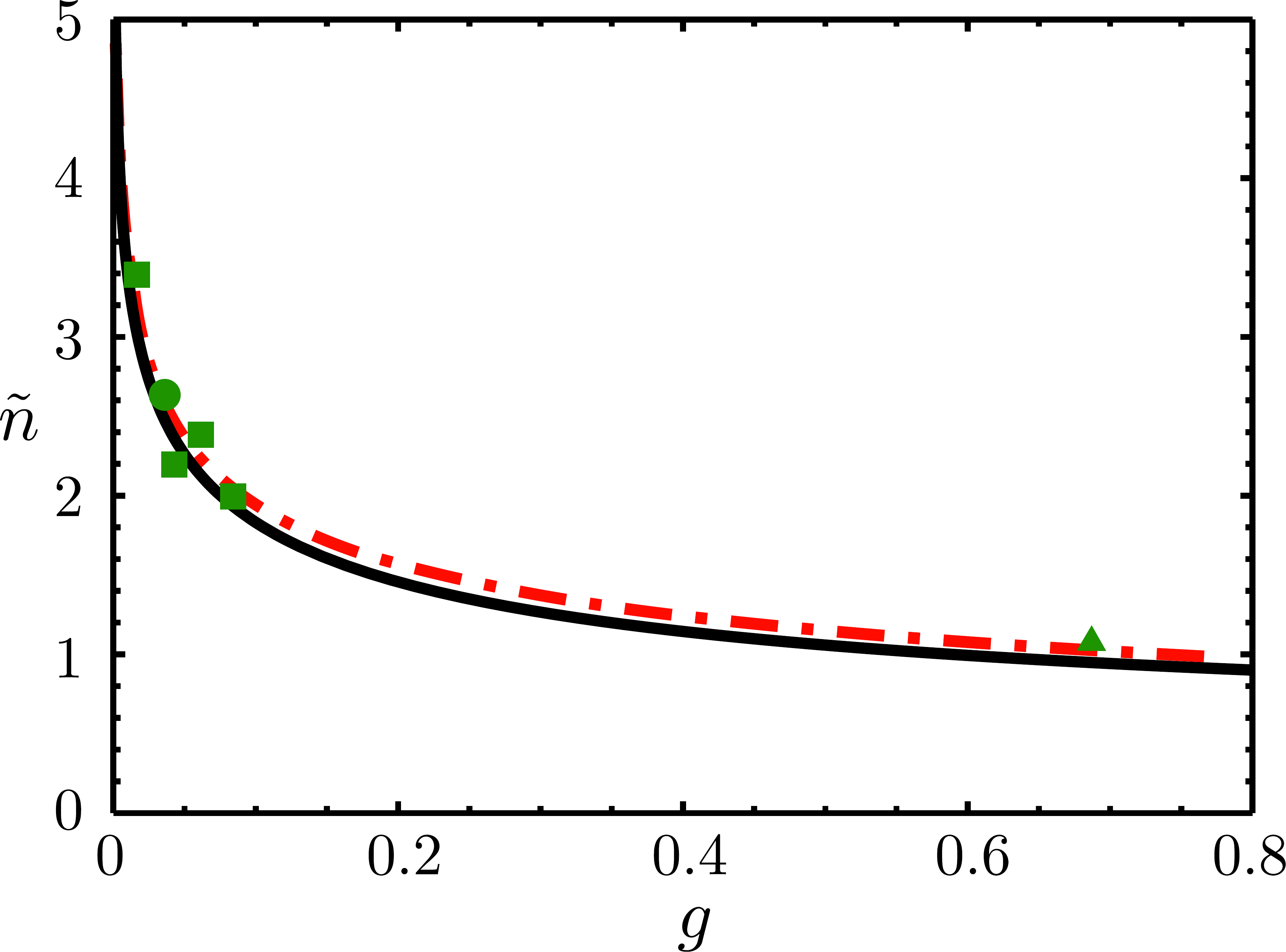}

%\vspace{2.5mm}

\includegraphics[width=0.975\linewidth]{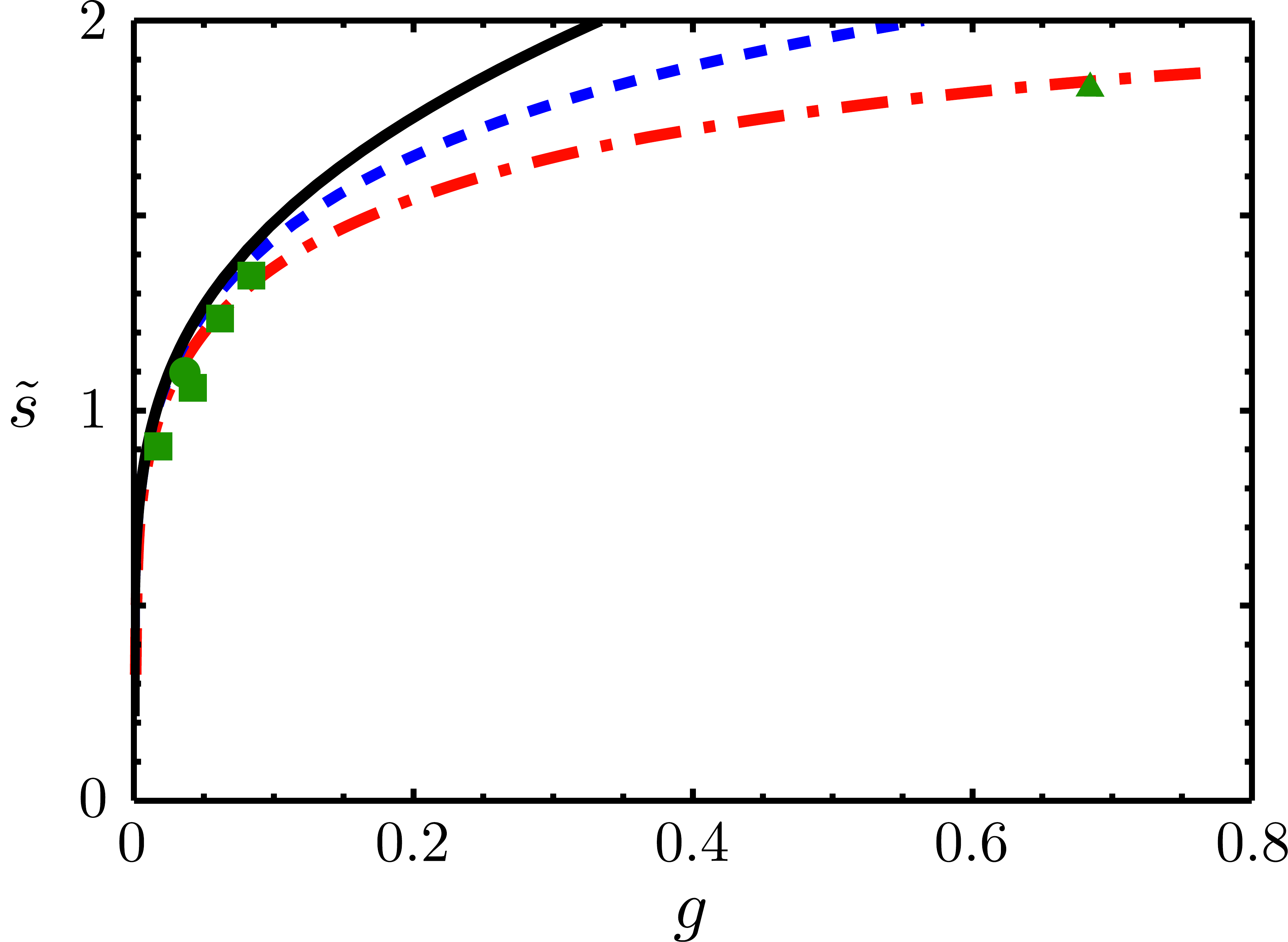}
\caption{
(Color online)
Comparison of our results for the renormalized state functions at $\mu = 0$ with the numerical results of Ran\c con and Dupuis\cite{Ran12} (dash-dotted red) as well as with data from three different experiments 
(green symbols, taken from Fig.~4 in [\onlinecite{Ran12}]). 
The remaining lines correspond to the full analytical expressions in Eqs.~(\ref{eq:result_pressure_free}) and (\ref{eq:result_entropy_free}) on the one hand (dashed blue), and to the improved
equations~(\ref{eq:pspress_better_result_long}), (\ref{eq:result_density_free}), and (\ref{eq:entropy_equation_from_flow}) on the other hand (solid black), where $r$ is always taken from Eq.~(\ref{eq:sigma_result_general}). From top to bottom we show the reduced pressure $\tilde{p}$, the phase-space density $\tilde{n}$, and the entropy per particle $\tilde{s}$ as a function of the effective coupling constant $g$ as
defined in Eq.~(\ref{eq:g_def}).
}
\label{fig:results_mu_0_with_numerics}
\end{figure}

In Fig.~\ref{fig:results_mu_0_with_numerics} we compare our results for the thermodynamic state functions at $\mu = 0$ (solid black and dashed blue lines) with empirical data from three ultra-cold atoms experiments (green symbols) which we call Chicago I,\cite{Hun11} ENS,\cite{Yef11} and Chicago II,\cite{Zha12} following the naming introduced in Ref.~[\onlinecite{Ran12}]. The first two experiments investigated $\ce{^{133}Cs}$ and $\ce{^{87}Rb}$ atoms, respectively, inside a harmonic potential with strong confinement along the $z$-axis, resulting in a quasi two-dimensional system; connection to homogeneous systems was made through the local density approximation. While the 3D scattering length in Ref.~[\onlinecite{Yef11}] was fixed at $a = 5.3$~nm ($g = 0.035$), the Chicago I experiment used a magnetic Feshbach resonance to vary $a$ between $2 - 10$~nm ($g = 0.016 - 0.083$). These values of $g$ correspond to $W(1/g) = 1.9 - 3.0$ which is not much larger than unity, in contrast to the assumption in our calculations; nevertheless we find that the agreement between the analytical results and the aforementioned experiments is quite good. On the other hand the measurements from the Chicago II experiment, based on $\ce{^{133}Cs}$ atoms in a two-dimensional optical lattice, differ visibly from our predictions for $\tilde{p}$ and $\tilde{s}$ which is not surprising given that $g = 0.68$ is relatively large, while the agreement for $\tilde{n}$ is still remarkably good.

As a further benchmark we use the results of Ran\c con and Dupuis\cite{Ran12} (dash-dotted red lines) who numerically computed the state functions within the FRG using a truncated gradient expansion. We find that the plots for the phase-space density $\tilde{n}$ essentially agree, while the reduced pressure $\tilde{p}$ and the entropy per particle $\tilde{s}$ differ for $g \gtrsim 0.1$. This indicates the upper boundary of the regime where our asymptotic analysis is valid for these state functions.

\section{Quantum Monte Carlo Simulations of the $XY$ model}
\label{sec:quantum_monte_carlo}
In this section we compare our analytic RG results for the density and the compressibility in the vicinity of the dilute Bose gas quantum critical point derived in  Sec.~\ref{sec:thermodynamics} with numerical results for the two-dimensional quantum $XY$ model in a magnetic field which we have investigated using QMC simulations with finite-size scaling on lattice sizes from $20 \times 20$ up to $100 \times 100$ spins. We have implemented the stochastic series expansion algorithm \cite{Sandvik2002} with directed loop updates and using the so-called Mersenne Twister random number generator.\cite{Mersenne1998} The Hamiltonian is given in terms of the components $\hat{S}^{\alpha}_i$ of spin-$1/2$ operators localized at the sites $i$ of a square lattice,
\begin{align}
\mathcal{H} = J \sum_{\langle i j \rangle} \left( \hat{S}^{x}_{i} \hat{S}^{x}_{j} +  \hat{S}^{y}_{i} \hat{S}^{y}_{j} \right) - B \sum_{i=1}^{N} \hat{S}_{i}^{z},
\label{eq:hamiltonian_xy_model}
\end{align}
where the first sum is over distinct pairs of nearest neighbors on the square lattice with $N=L\times L$ lattice sites and periodic boundary conditions in both directions. Here $J$ is the nearest neighbor exchange coupling and the magnetic field $B$ is measured in units of energy. The model (\ref{eq:hamiltonian_xy_model}) maps exactly to the two-dimensional Bose-Hubbard model with infinite onsite interaction,\cite{Sac11} i.e., hard-core bosons. To compare both models, we should therefore take the limit of infinite contact interaction ($f_0 \rightarrow \infty$) in our boson Hamiltonian (\ref{eq:hamiltonian}) so that only the logarithmic term in Eq.~(\ref{eq:g_def}) survives,
\begin{align}
g (u_0 \to \infty) = \frac{4}{\ln \left[ \frac{\Lambda_0^2}{2 m (T - \mu)} \right]}.
\end{align}
This limit makes it more challenging to reach the regime $g \ll 1$ where our theoretical results apply as this requires exponentially low temperatures, in contrast to cold atoms experiments where the bare interaction $u_0$ can be tuned to small values (cf. the discussion of Fig.~\ref{fig:results_mu_0_with_numerics}). Therefore, numerical simulations on hard-core bosons are suitable to test the limitation of our approximations as we will see below. The quantum critical points of the $XY$ model at $B  = \pm 2 J$ belong to the same universality class as the dilute Bose gas in Eq.~(\ref{eq:hamiltonian}). Moreover, at the critical fields $B = \pm 2 J$ the bare parameters of the above $XY$ Hamiltonian (\ref{eq:hamiltonian_xy_model}) can be related to the bare mass and the chemical potential of the dilute Bose gas via $m  = 1/( J a^2) $ and  $\mu = 2J \mp B$. The magnetization per site $M/N$ in the simulations is directly related to the boson density $n = {M}/{N} \mp {1}/{2}$ and the longitudinal spin-susceptibility $\chi$ of the $XY$ model corresponds to the compressibility $\kappa$ of the dilute Bose gas. In what follows we set the lattice spacing $a$ to unity.

\subsection{Results for $\mu = 0$}
\label{sec:QMCmu0}

\begin{figure} %[t]
  \centering
   \includegraphics[width=\columnwidth]{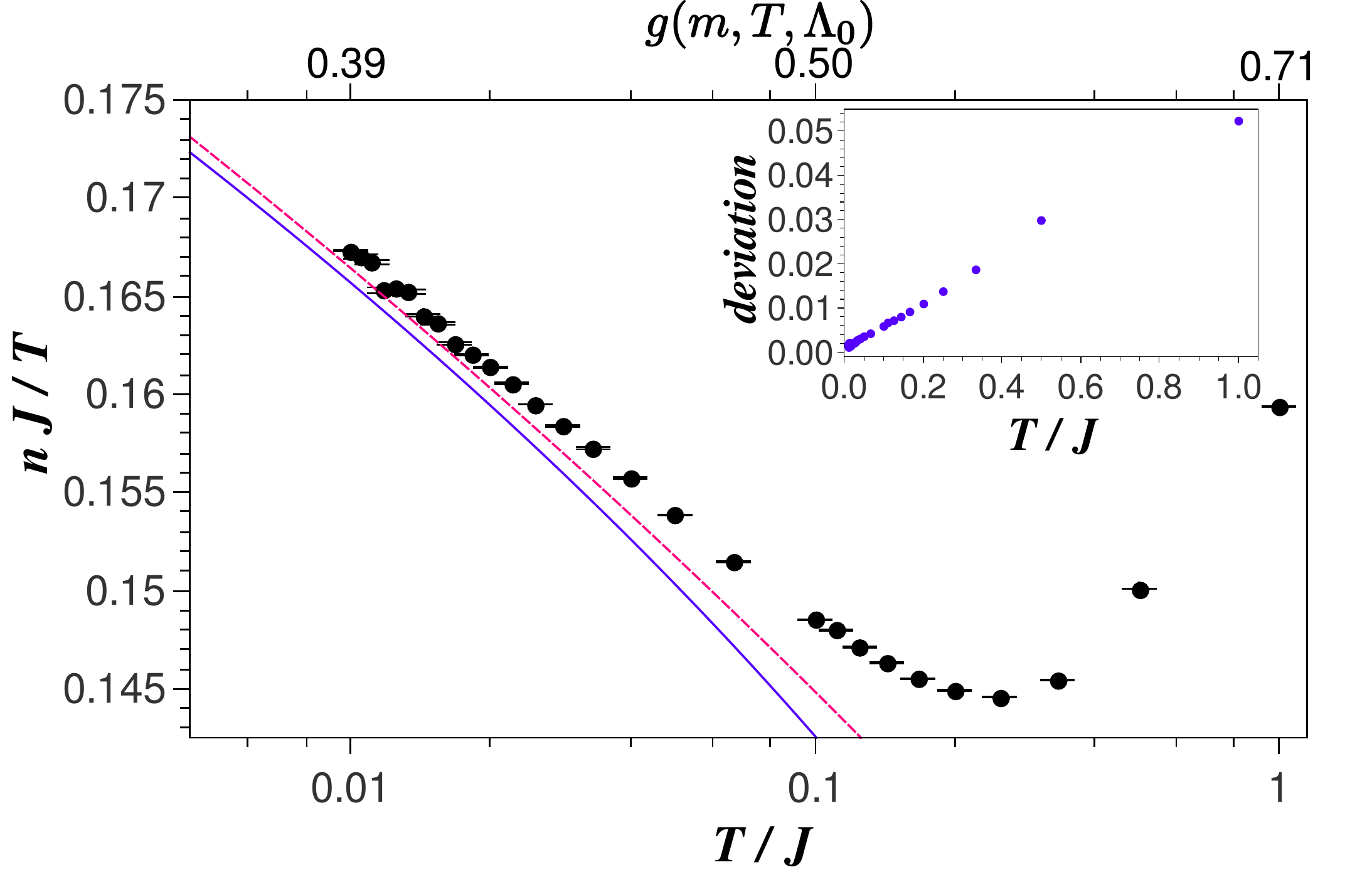}
  \vspace{-3mm}
  \caption{(Color online) Density over temperature as a function of $T$ for the $XY$ model at $\mu = 0$ from QMC simulations (black dots). The blue line represents our analytical prediction in Eqs.~(\ref{eq:density_gen_result_2D}) and \eqref{eq:sigma_result_general_posMu} with $m J=1$ and $\Lambda_0  = 23.5$. The dashed line (red) is a fit of the QMC data to our leading order result~(\ref{simplelog}) for the density using a larger cutoff $\Lambda_0 = 205$ and a modified mass $m J = 0.8$. On the top axis we show the corresponding $g$ values for $m J=1$ and $\Lambda_0  = 23.5$.}
\label{fig:QMC-mu0}
\end{figure}

The magnetization data from the simulations follows a characteristic finite-size scaling of the form $M(L)/N = M(\infty)/N + b\, \exp(-L/\xi')$, where $b$ and $\xi'$ are temperature 
dependent as will be discussed in more detail in Sec.~\ref{sec:QMC-CorrelationLength-Validity}. Basically the finite-size correlation length $\xi'$ decreases with increasing temperature 
up to some temperature $T_{\rm end}$, above which the data becomes 
largely independent of $L$ for the system sizes used. In Fig.~\ref{fig:QMC-mu0} we show the QMC results with error bars (black) for the density over temperature as a function of $T$ in the thermodynamic limit  at the lower critical field $B = -2J$, corresponding to a vanishing chemical potential $\mu = 0$. The solid line (blue) is a fit using Eq.~(\ref{eq:density_gen_result_2D}) together with Eq.~(\ref{eq:sigma_result_general_posMu}) in the limit $1/u_0\to 0$ for hard-core bosons. Keeping both $m$ and $\Lambda_0$ as fitting parameters we obtain the expected result $m J= 1$ within a few percent. Setting $mJ = 1$ for simplicity and using only the ultraviolet cutoff $\Lambda_0$ as a fitting parameter, we obtain $\Lambda_0  = 23.5 \pm 1.5$ where we have required that the deviation between our analytical result and the QMC data vanishes in the limit $T \to 0$ (see inset of Fig.~\ref{fig:QMC-mu0}).

\begin{figure} %[t]
  \centering
\includegraphics[width=\columnwidth]{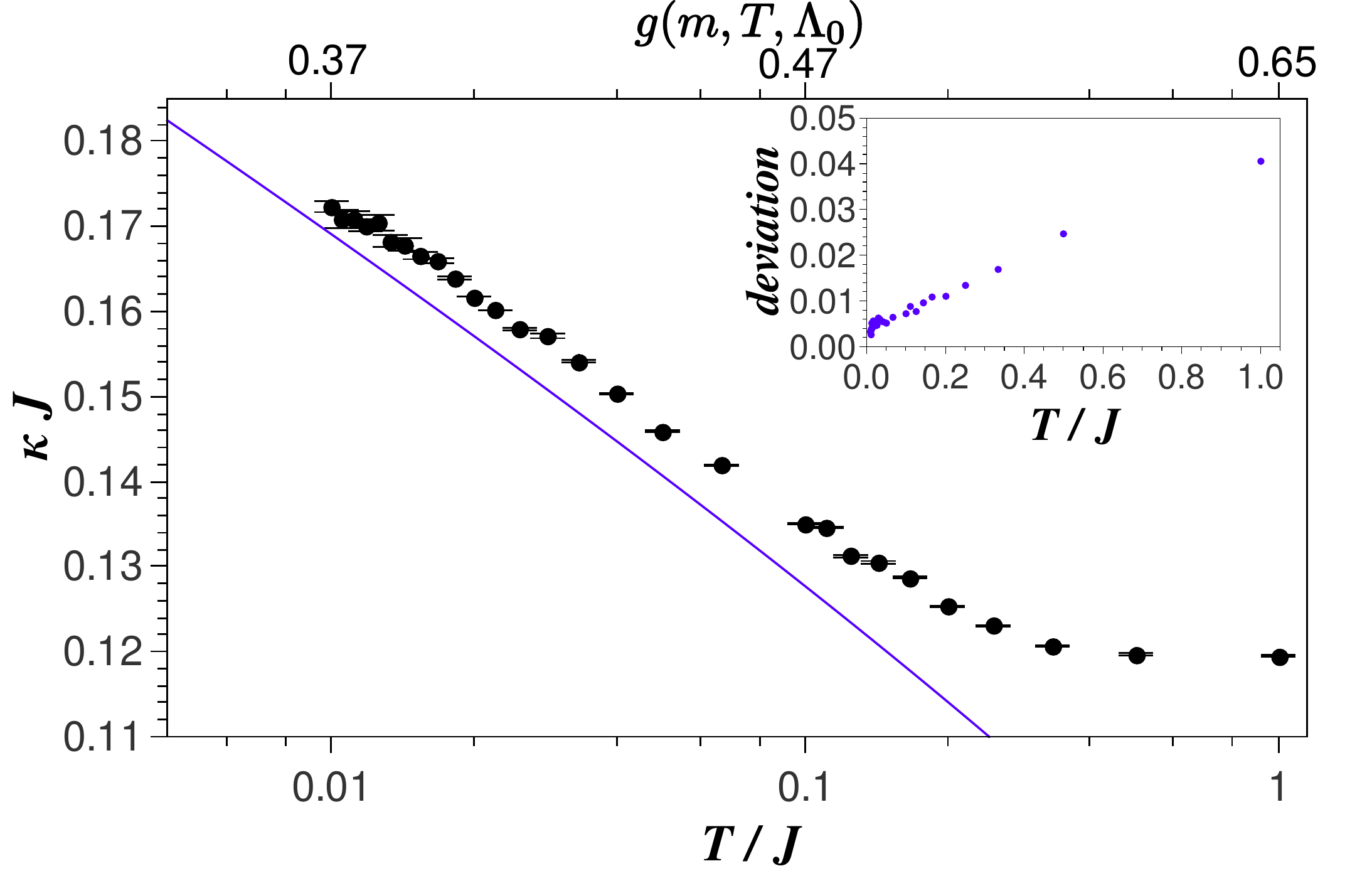}
  \vspace{-3mm}
  \caption{(Color online) Compressibility as a function of temperature for the $XY$ model at $\mu = 0$ from QMC simulations (black dots). For comparison we show as a blue line 
our analytical prediction from Eqs.~(\ref{eq:density_gen_result_2D}),
(\ref{eq:sigma_result_general_posMu}), and Eq.~(\ref{eq:kappa}) with $mJ=1$ 
and $\Lambda_0  = 31 $. The top axis shows the corresponding $g$ values for these parameters.}
\label{fig:kappa}
\end{figure}

For comparison we also tested the leading order expression for the density given in Eq.~(\ref{simplelog}) using the same cutoff $\Lambda_0 = 23.5$, which predicts $n/T$ to be more than $40\%$ below the QMC data for the temperatures used here. It is possible to use a rather different cutoff $\Lambda_0 \approx 205$ and a modified mass $mJ \approx 0.8$ to fit the leading order expression in Eq.~(\ref{simplelog}) to the QMC data as shown by the dashed line (red) in Fig.~\ref{fig:QMC-mu0}. It seems that such a fit compensates higher order corrections by using a modified effective mass and a large value of the cutoff. In turn this means that for most experimental and numerical data the coupling constant $g$ is sufficiently small to guarantee that the approximate solution in Eq.~(\ref{eq:density_gen_result_2D}) is accurate, but $g$ is not exponentially small to justify dropping all higher order terms (see $g$-axes in Fig.~\ref{fig:QMC-mu0} and Fig.~\ref{fig:kappa}). Therefore, a fit to the simple logarithmic behavior in Eq.~(\ref{simplelog}) may yield an incorrect cutoff and mass to compensate different higher order corrections. In the temperature region around the minimum in Fig.~\ref{fig:QMC-mu0} the scaling of the density may even appear perfectly linear with temperature, which is consistent with recent results on two-dimensional coupled spin-dimers systems.\cite{St15}

The corresponding data for the compressibility is shown in Fig.~\ref{fig:kappa}. The same form of the finite-size scaling was used, but it should be noted that convergence to the thermodynamic limit requires larger system sizes for this response function. Using Eq.~(\ref{eq:kappa}) and the analytic expression in Eq.~(\ref{eq:kappa-leading}) we find good agreement using again $mJ=1$, but a larger cutoff $\Lambda_{0}  = 31 \pm 2$. Note that the effective coupling $g$ in Eq.~(\ref{eq:g_def}) depends only logarithmically
on $\Lambda_0$ so that the difference in the cutoffs only results in a rather small correction. Nevertheless we find that we cannot choose a single cutoff to fit our analytical results for both the density and the compressibility to the QMC data. We have checked the size-scaling for the particle density as well as for the compressibility very carefully, so that we can be sure that the deviation in the cutoff is not related to 
any remaining finite-size effects. Instead, the different values of $\Lambda_0$ might be due to the fact that the condition $g \ll 1$, requiring exponentially low 
temperatures due to the hard-core interaction, is not strictly fulfilled in the accessible temperature regime (see Fig.~\ref{fig:QMC-mu0} and Fig.~\ref{fig:kappa}). Hence we would expect that our analytical result (\ref{eq:sigma_result_general_posMu}) is only qualitatively correct and we need different values of the cutoff for quantitative agreement with the QMC simulations.

\begin{figure} %[t]
  \centering
  \includegraphics[width=\columnwidth]{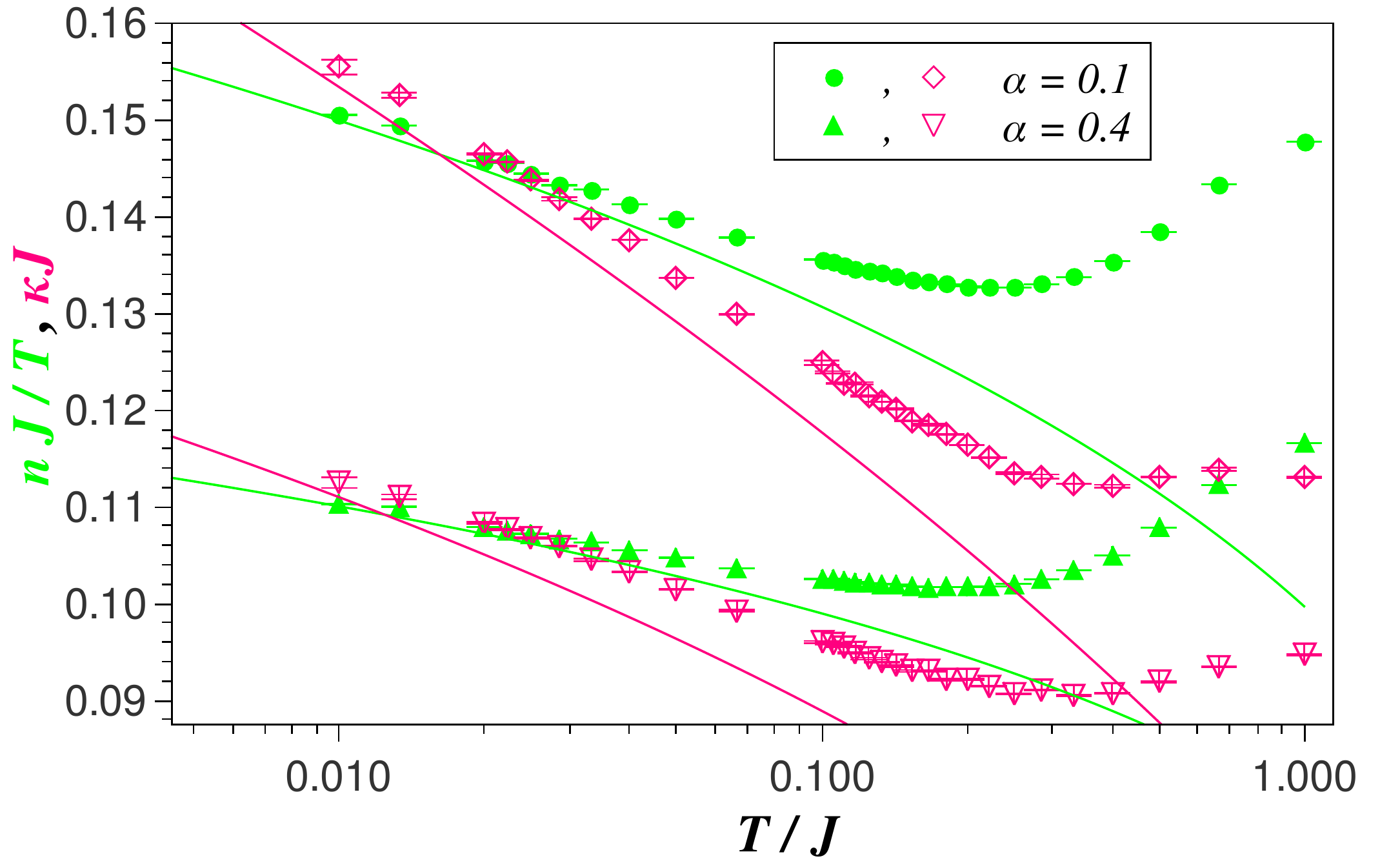}
  \vspace{-3mm}
  \caption{(Color online) QMC results for the $XY$ model at $\alpha = 0.1$ and $\alpha = 0.4$ of the density over temperature (green solid symbols) and the compressibility (red open symbols). The green solid lines correspond to our analytical prediction for $n / T$ from Eqs.~\eqref{eq:density_gen_result_2D} and \eqref{eq:sigma_result_general_posMu} with $mJ = 1$, $\Lambda_0  = 23 \pm 2$ (upper green line), and $\Lambda_0  = 21 \pm 2$ (lower green line), while the red solid lines correspond to $\kappa$ from Eq.~\eqref{eq:kappa} with $mJ = 1$, $\Lambda_0  = 27.5 \pm 2$ (upper red line), and $\Lambda_0  = 20 \pm 2$ (lower red line).}
\label{fig:QMC-alpha01+04}
\end{figure} 

\subsection{Results for $\mu \neq 0$}
\label{sec:QMCalpha01}
Next, let us consider the regime of constant, non-zero $\alpha=-\mu/T > 0$, corresponding to approaching the QCP diagonally from the left in the $\mu$-$T$-diagram shown in Fig.~\ref{fig:phase_diagram}. We investigate a wide range of values for $\alpha$, starting from $0.1$ up to $2.0$.

The QMC data (solid and open symbols) for the density and the compressibility at $\alpha = 0.1$ are shown in Fig.~\ref{fig:QMC-alpha01+04} as a function of temperature, where we compare them to our analytical results in Eqs.~\eqref{eq:density_gen_result_2D} and \eqref{eq:kappa} (solid lines).
The fit for the density shows good agreement for $mJ = 1$ and $\Lambda_0 = 23 \pm 2$ (upper green line), which is consistent with the estimate for $\mu = 0$ from above. For the compressibility we find good agreement using $mJ = 1$ and $\Lambda_0 = 27.5 \pm 2$ (upper red line). It should be noted that it is again possible to fit the data to a leading order expansion of our analytical results, but this leads to values for the cutoff which are even larger than in the case of $\alpha=0$.

Increasing $\alpha$ to $0.4$ as shown in Fig.~\ref{fig:QMC-alpha01+04} (lower graphs), we see that the magnitude of the density and the compressibility is lowered, but the overall shape is similar. Compared to the case $\alpha = 0$ the value of the respective cutoffs decreases, $\Lambda_0 = 21 \pm 2$ for the density and $\Lambda_0 = 20 \pm 2$ for the compressibility at $\alpha = 0.4$, while at the same time the difference between the two cutoffs becomes negligible within errors.

\begin{figure} %[t]
  \centering
  \includegraphics[width=\columnwidth]{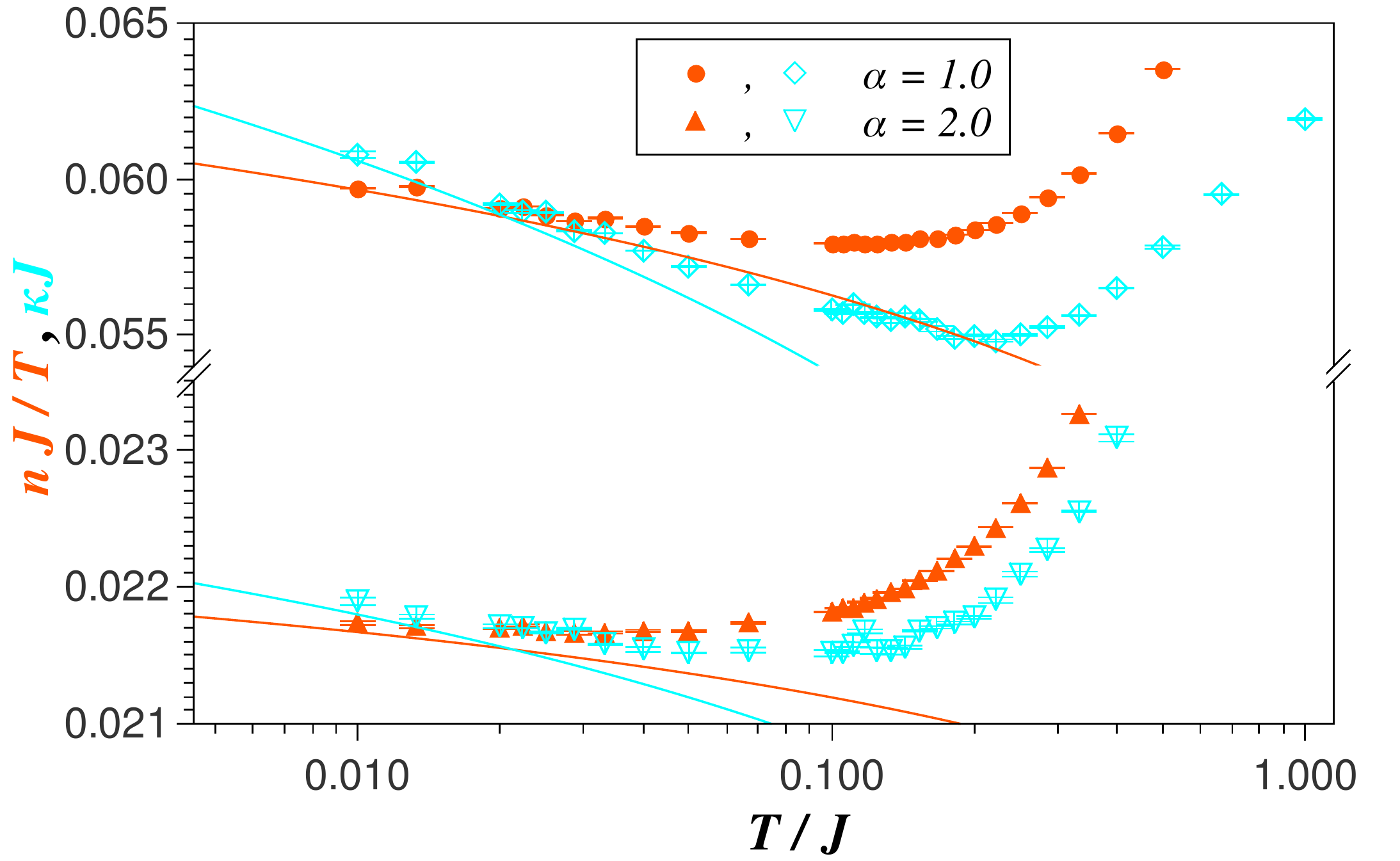}
  \vspace{-3mm}
    \caption{(Color online) QMC results for the $XY$ model at $\alpha = 1.0$ and $\alpha = 2.0$ of the density over temperature (orange solid symbols) and the compressibility (blue open symbols). The orange solid lines correspond to our analytical prediction for $n / T$ from Eqs.~\eqref{eq:density_gen_result_2D} and \eqref{eq:sigma_result_general_posMu} with $mJ = 1$ and $\Lambda_0  = 21 \pm 2$, while the blue solid lines correspond to $\kappa$ from Eq.~\eqref{eq:kappa} with $mJ = 1$ and $\Lambda_0  = 20 \pm 2$.}
\label{fig:QMC-alpha1+2}
\end{figure} 

Monte Carlo results for $\alpha = 1$ and $\alpha = 2$ are presented in Fig.~\ref{fig:QMC-alpha1+2}; with increasing $\alpha$, the temperature region for which we can apply the fit functions is pushed to lower and lower values of $T$, while at the same time the characteristic minimum in $n/T$ at $T_{\text{min}}$ is shifted to lower temperatures as well. Note that $\kappa$ also shows 
a minimum at slightly larger temperatures which does not shift as much with $\alpha$, so that both minima approach each other for $\alpha = 2$. Remarkably, both $\kappa$ and $n/T$ take on the same value at a certain crossing temperature of $T = 0.0228 \pm 0.0018$ (QMC) or $T = 0.018 \pm 0.003$ (fits), which is largely independent of $\alpha$.

From the fits of the density the cutoff can be consistently 
estimated to be in the range  
$\Lambda_0 \approx 22 \pm 2$ for all values of $\alpha$.   
The data for the compressibility at small $\alpha$ gives slightly larger estimates
for $\Lambda_0$, indicating that in this case our analytical results for $\kappa$ require smaller values of $g$ than for the density. This is consistent with Fig.~\ref{fig:results_mu_0_with_numerics} where we find good agreement with Ref.~[\onlinecite{Ran12}] for $n/T$ in a surprisingly large regime of $g$, while our results for the other observables only agree up to $g \lesssim 0.1$.

\subsection{Correlation length and validity range}
\label{sec:QMC-CorrelationLength-Validity}
\begin{figure} %[t]
  \centering
  \includegraphics[width=\columnwidth]{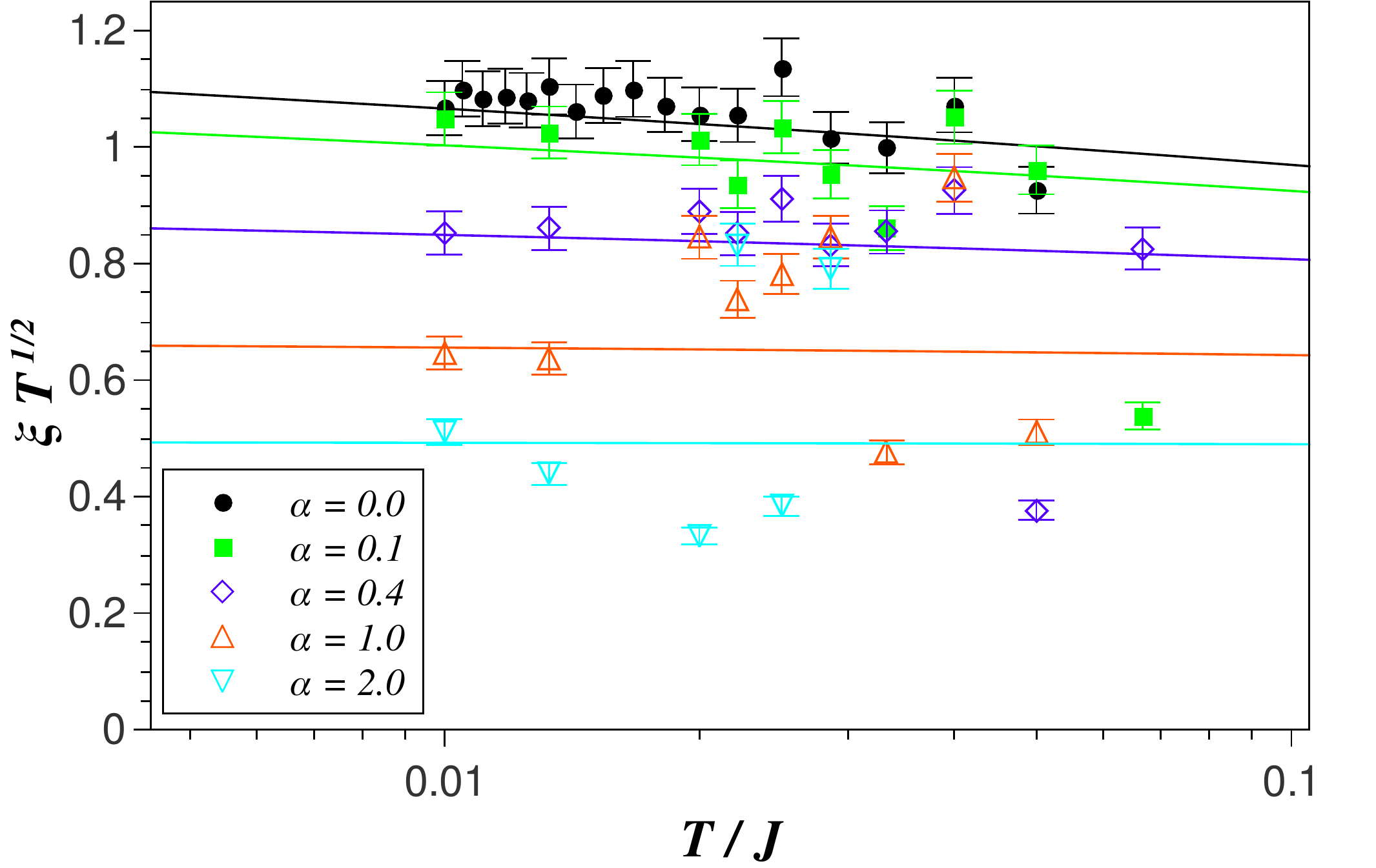}
  \vspace{-3mm}
  \caption{(Color online) Comparison of the numerical results for the rescaled finite-size correlation length $\sqrt{T} \xi = 1.15 \sqrt{T} \xi'$ (symbols) at different values of $\alpha = -\mu/T$ to our analytical prediction $\sqrt{T} \xi = 1 / \sqrt{2 m r}$ from Eqs.~\eqref{eq:correlationlength} and \eqref{eq:sigma_result_general_posMu} (solid lines) with $mJ = 1$ and using a cutoff of $\Lambda_0 = 22$.}
\label{fig:CorrelationLength}
\end{figure}

As already mentioned before, the size scaling of the magnetization and of the susceptibility has the form
\begin{align}
X (L) = X (\infty) + b\, \exp(-L / \xi')
\label{expfit}
\end{align}
up to a characteristic temperature $T_{\text{end}}$, above which the data is mainly independent of $L$ so that a simple linear extrapolation suffices. With these fits we can extract a finite-size correlation length $\xi'$ from the size scaling of the magnetization, which is shown in Fig.~\ref{fig:CorrelationLength} for different $\alpha$ as a function of temperature using the rescaling $\xi=1.15 \xi'$. Within error bars the behavior of $\xi'$ is consistent with the divergence $\xi \propto 1/\sqrt{T}$ in Eq.~(\ref{eq:correlationlength}) for all $\alpha$. For comparison we show the analytic result from Eq.~(\ref{eq:sigma_result_general_posMu}) for $\sqrt{T} \xi = 1/\sqrt{2 m r}$ as solid lines for different values of $\alpha$.  The correlation length $\xi'$ from the finite-size scaling of the Monte Carlo data discussed above shows a similar scaling behavior, although finite-size scaling does not measure the correlation length directly. Choosing the cutoff as $\Lambda_0 = 22$ consistent with the fits for the density, we observe that the correlation length $\xi'$ from QMC finite-size scaling agrees reasonably well with the analytic prediction for $\xi$ up to the rescaling $\xi =1.15 \xi'$.

\begin{figure} %[t]
  \centering
  \includegraphics[width=\columnwidth]{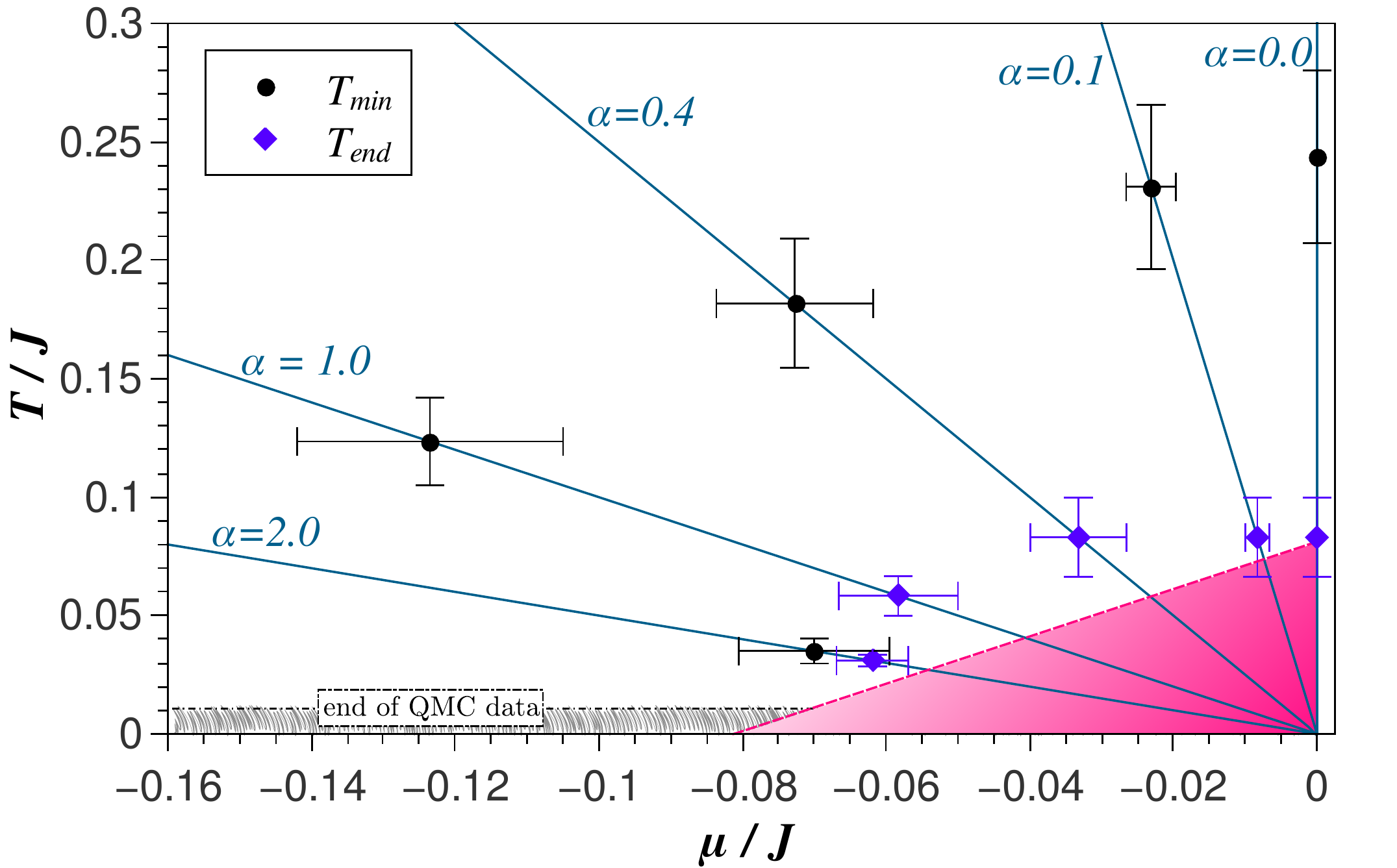}
  \vspace{-3mm}
  \caption{(Color online) Region of validity (red area) where the extrapolation in Eq.~(\ref{expfit}) works and our analytical predictions from the FRG yield accurate results. For comparison we also show the temperature $T_{\rm min}$ where $n/T$ exhibits a minimum for a given $\alpha$. The dashed red line is given by $g=0.5$ for $mJ=1$ and $\Lambda_0 = 22$.}
\label{fig:ValidityRange}
\end{figure}

The breakdown of the exponential extrapolation in Eq.~(\ref{expfit}) defines a temperature $T_{\rm end}$, which can be used to identify a region of validity for the continuum description of the $XY$ lattice model. Above this temperature the correlation length $\xi$ is of order unity so that lattice effects dominate. As can be seen in Fig.~\ref{fig:ValidityRange}, this region where the continuum description holds roughly coincides with the parameter range $g < 0.5$ for $\Lambda_0=22$ (red shaded region), which is where our analytic prediction for $n$ in (\ref{eq:density_gen_result_2D}) can be fitted to the QMC data. The values of the temperatures $T_{\rm min}$ where the minimum in $n/T$ for a given $\alpha$ occurs are also shown and are generally at significantly larger values. Note that these minima in $n/T$ correspond to linear behavior of the density with temperature, which always occurs well above the region of validity. It is remarkable that it is possible to use Eq.~(\ref{eq:density_gen_result_2D}) to describe the behavior close to the QCP also for finite $\mu$ rather accurately, but as expected smaller temperatures are required for $\alpha > 0$ since we then approach the QCP diagonally in the $\mu$-$T$-diagram.

\section{Quasi-particle properties close to the quantum critical point}
\label{sec:quasiparticle_properties}

Within our FRG approach, it is straightforward to calculate the
leading order momentum and frequency dependence of the  self-energy in the vicinity of the
dilute Bose gas quantum critical point, which we parametrize
in terms of the two dimensionless renormalization factors $Z_{\Lambda}$ and $Y_{\Lambda}$
defined via the expansion (\ref{eq:self_energy_expansion}).
$Z_{\Lambda}$ can be identified with the usual wavefunction renormalization factor
(quasi-particle residue),
while $Y_{\Lambda}$ determines the effective mass of the bosons.

\subsection{Wavefunction renormalization}
\label{sec:wavefunction_renormalization}
\begin{figure}
\includegraphics[width=\linewidth]{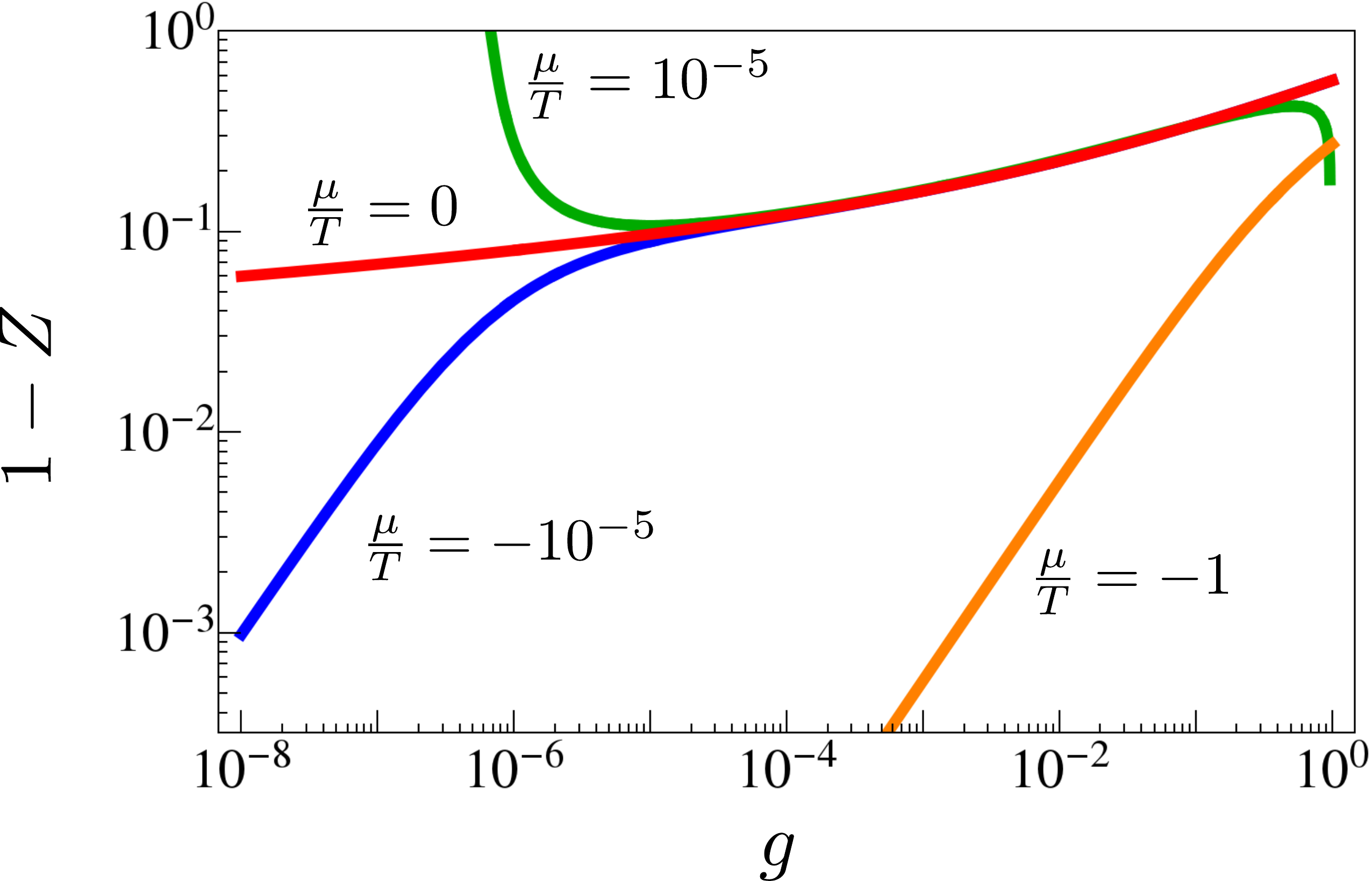}
\caption{
(Color online)
Double logarithmic plot of our analytical result for $1 - Z$ 
as given in Eq.~(\ref{eq:z_factor_asymptotic}) versus  the dimensionless coupling 
$g$ for different values of $\mu / T$.
For $\mu / T = \pm 10^{-5}$ we can see that the scaling is very close to the scaling at 
vanishing chemical potential for larger $g$ and only begins to differ when $g W(1/g)$ is of the order of $|\mu / T|$;
the deviation around $g \approx 1$ is due to the fact that for positive chemical potential we have to use (\ref{eq:sigma_result_general_posMu}) for $r$ instead of (\ref{eq:sigma_result_general}). 
In the case of positive $\mu$ the wavefunction renormalization then shrinks until it vanishes at the phase transition where $r = 0$ [see Eq.~(\ref{eq:phase_transition_g})].
}
\label{fig:z_factor_absolute}
\end{figure}

From the low-energy expansion  (\ref{eq:self_energy_expansion}) 
of the self-energy we see that
the RG flow of $Z_\Lambda$ is determined by
\begin{equation}
\partial_\Lambda Z_\Lambda^{-1} = - \partial_\Lambda \partial_{\omega} \Sigma_\Lambda (0, \omega + i 0^+) \bigr|_{\omega = 0}.
\end{equation}
Although the flow of $Z_{\Lambda}$ modifies also the flow equations 
(\ref{eq:flow_equation_r_self_energy}) 
and (\ref{eq:flow_equation_pi_self_energy})
for $\Sigma_\Lambda (0)$ and $\Pi_\Lambda (0)$, we shall ignore this modification since 
it is not expected to change the leading asymptotics close to the quantum critical point.
Neglecting the Bose distribution in the flow of the particle-particle susceptibility 
we end up with the approximate flow equation (see appendix~\ref{sec:detailed_z_factor})
\begin{equation}
\partial_l Z_l^{-1} = \frac{2 g \beta_l e^{\beta_l + r}}{ \left( e^{\beta_l + r} - 1 \right)^2}.
\end{equation}
Integrating this equation we find that
$Z = \lim_{ l \rightarrow \infty} Z_l$ is given by
\begin{equation}
Z^{-1} - 1 \approx \frac{g}{e^r - 1}.
\label{eq:z_factor_asymptotic}
\end{equation}
Although this result is only strictly valid for non-positive chemical potential, we find numerically that it is also qualitatively good for $\mu > 0$. For vanishing chemical potential $Z$ scales as
\begin{equation}
Z^{-1} - 1 = \frac{1}{W(1/g)},
\end{equation}
while in the quantum disordered regime $\alpha \gg 1$ the correction is exponentially small,
\begin{equation}
Z^{-1} - 1 = g e^{-\alpha}.
\end{equation}
We have verified the validity of these approximate expressions 
by solving the relevant flow equations numerically, 
taking the Bose distribution in the flow of $\Pi_\Lambda$ into account, 
which yields very good overall agreement. Some representative results
for  $1-Z$ at various values of $\mu / T$ are presented in Fig.~\ref{fig:z_factor_absolute}. We find that for $\mu / T \leq 0$ the wavefunction renormalization approaches unity in the limit $g \to 0$ in agreement with the well-known result that $\Sigma(K) = 0$ at the QCP (cf. Sec.~\ref{sec:exact_at_qcp}), while for positive chemical potential we have $Z = 0$ at a finite $g$ where the BKT transition takes place.

Finally, we can also explore the regime of validity of the
self-consistent $T$-matrix approximation as given in Eqs.~(\ref{eq:self_consistent_t_matrix_sigma}) and (\ref{eq:self_consistent_t_matrix_pi}), which in this context is expected to be good at high temperatures. On the other hand, our FRG approach should be accurate at low temperatures
and it is a priori not clear whether there exists an intermediate temperature regime
where both methods are valid.
For simplicity we consider only the case $\mu =0$ and take the limit of infinite bare
interaction, corresponding to hard-core bosons.
A detailed discussion of the self-consistent $T$-matrix  approach to hard-core bosons
can be found in Ref.~[\onlinecite{Str15}], where the  
spin Hamiltonian for the magnetic insulator
 $\text{Cs}_2\text{CuCl}_4$ was mapped onto a two-dimensional hard-core boson model
which was then studied using the  self-consistent $T$-matrix  approximation.

\begin{figure}
\includegraphics[width=\linewidth]{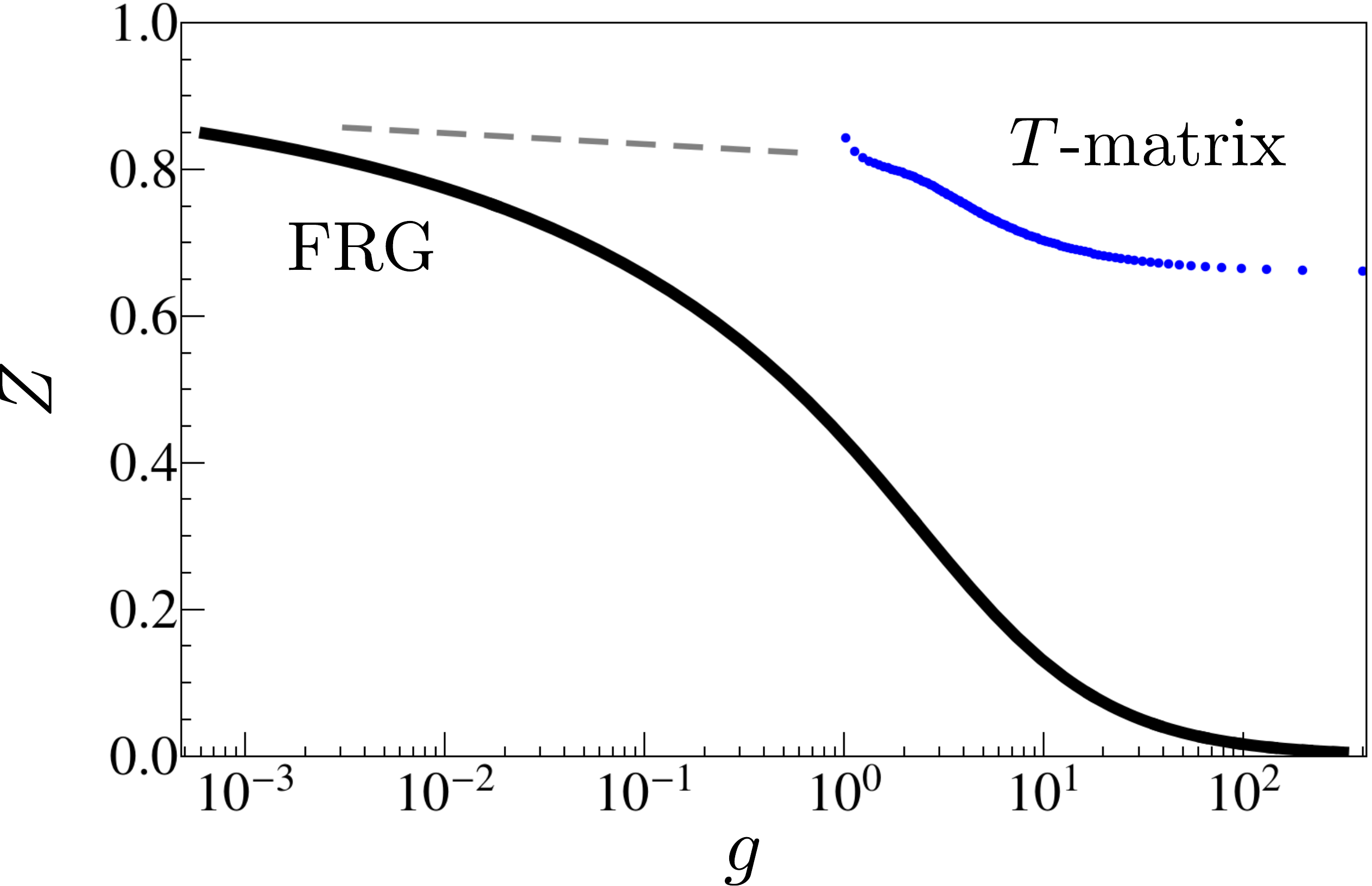}
\caption{
(Color online)
Comparison of our analytical result for the wavefunction renormalization  
$Z$ of hard-core bosons in Eq.~(\ref{eq:z_factor_asymptotic}) 
 (black solid line) 
with numerical computations using the self-consistent $T$-matrix approximation which is expected to be good at high temperatures (blue dots). 
Note that for $T \rightarrow 0$ our analytic result for $Z$ becomes exact.
The intermediate regime (sketched by the gray dashed line as a simple interpolation between the results) where $g = (\frac{1}{4} \ln \beta_0)^{-1}$ is below, but not much smaller than unity, is not covered by either method.
}
\label{fig:z_factor_interpolation_cs2cucl4}
\end{figure}

For a comparison of this method with our FRG approach
we use  parameters specific to $\text{Cs}_2\text{CuCl}_4$; in particular, we choose the effective inverse temperature as
 $\beta_0 = \Lambda_0^2/(2 m T) \approx 1 \text{K} / T$,
where we have fixed
the momentum cutoff $\Lambda_0$ to an average value of the inverse lattice 
parameters of $\text{Cs}_2\text{CuCl}_4$.
In Fig.~\ref{fig:z_factor_interpolation_cs2cucl4} we compare the results of
both complementary methods; obviously, an 
intermediate temperature regime where both methods are accurate does not exist,
showing the need for an alternative approach in this region. We expect that a numerical solution of the FRG flow equations \eqref{eq:flowselfdilute} and \eqref{eq:flowpoldilute}, retaining the full momentum and frequency dependence of $\Sigma_\Lambda (K)$ and $\Pi_\Lambda (P)$, should be accurate at low as well as at high temperatures. Another possibility to describe the intermediate temperature regime is to use a lattice FRG scheme along the lines of Ref.~[\onlinecite{Ran14}].


\subsection{Effective mass}
\begin{figure}
\includegraphics[width=\linewidth]{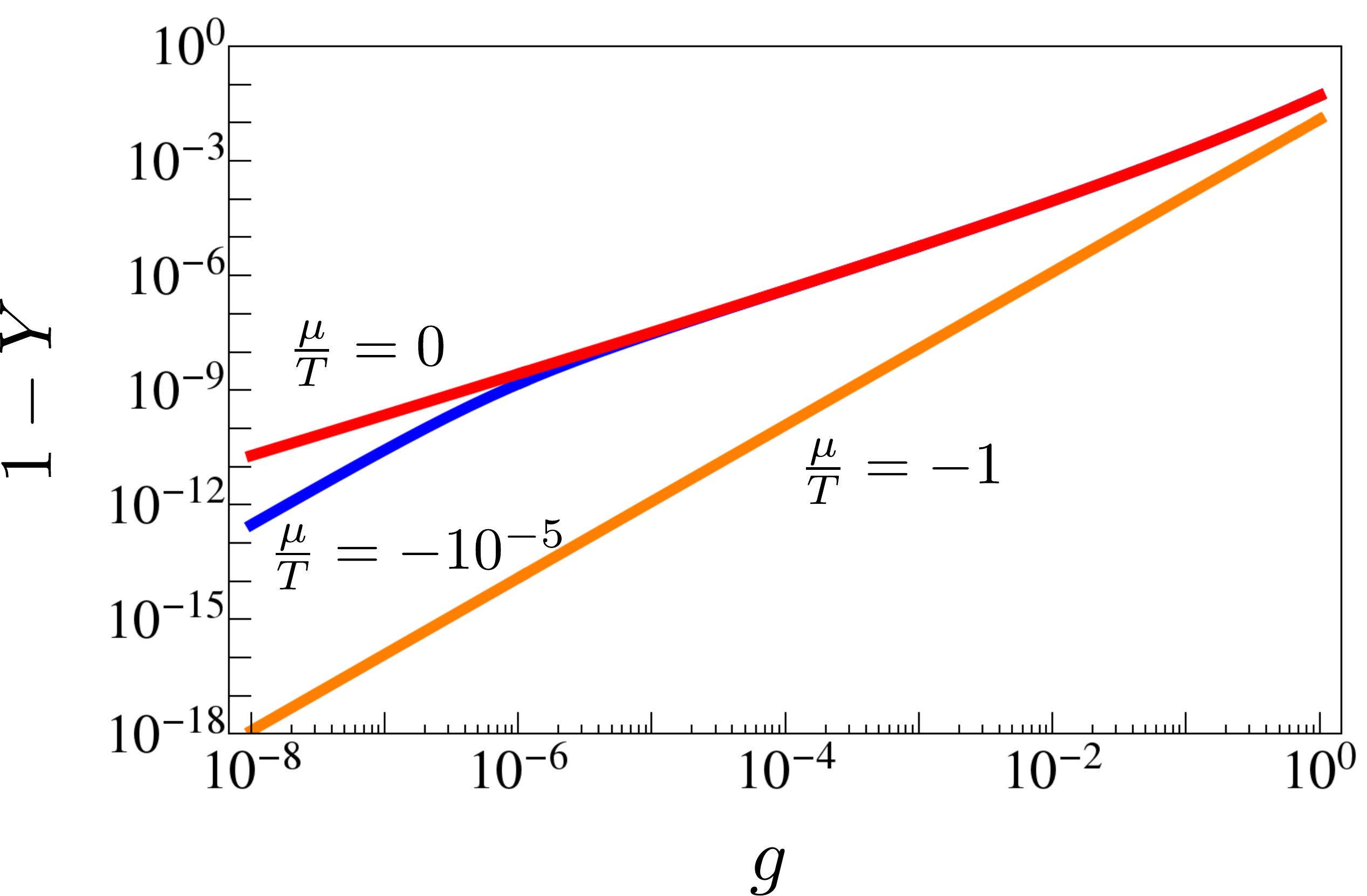}
\caption{
(Color online)
Double logarithmic plot of our analytical result 
for $1 - Y$ as given in Eqs.~(\ref{eq:y_small_r}) and (\ref{eq:y_large_r}) versus the dimensionless 
coupling $g$ for different values of $\mu /T$. 
We can see that for $\mu / T  = -10^{-5}$ the scaling coincides with the $\mu = 0$ curve until $g W(1/g)$ is of the order of $|\mu / T|$, where it starts to fall off more rapidly.
}
\label{fig:eff_mass_absolute}
\end{figure}

Using Eqs.~(\ref{eq:self_energy_expansion}) and (\ref{eq:flowselfdilute})
we find that the flow equation for the effective mass factor $Y_\Lambda$
is within our truncation given by
\begin{align}
\partial_\Lambda Y_\Lambda^{-1} &= m \partial_\Lambda \partial_k^2 \Sigma_\Lambda (\bm{k}, 0) \Big|_{k = 0}
\nonumber
\\*
& \hspace{-10mm} = - m \int_P \dot{G}_\Lambda (P) \biggl\{ 2 F_\Lambda^3 (P) \left[ \partial_k \Pi_\Lambda (P+K) \right]^2
\nonumber
\\*
&\hspace{18mm}- F_\Lambda^2 (P) \partial_k^2 \Pi_\Lambda (P+K) \biggr\} \bigg|_{K=0}.
\end{align}
As before we neglect the momentum dependence 
in the particle-particle susceptibility as well as the Bose distributions which appear in 
its flow equation. This allows us to compute $Y$ analytically to leading order as long as 
$\mu \leq 0$. For $r \ll 1$ we find
\begin{equation}
Y^{-1} - 1 = \frac{g^2}{32 r^2} \left[ g \ln (1/r) + \alpha \right],
\label{eq:y_small_r}
\end{equation}
which for $\mu = 0$ and small $g$ simplifies to
\begin{equation}
Y^{-1} - 1 = \frac{g}{32 W(1/g)}.
\end{equation}
In the opposite regime $r \gtrsim 1$ we obtain
\begin{equation}
Y^{-1} - 1 = \gamma (\alpha) g^2,
\label{eq:y_large_r}
\end{equation}
where the coefficient $\gamma$ is given by
\begin{equation}
\gamma (\alpha) = \frac{\alpha}{16} \int_0^{\infty} \frac{db}{(e^{b + \alpha} - 1) (b + \alpha)^2}.
\end{equation}
Thus at $\alpha \gg 1$ the correction is again exponentially suppressed,
\begin{equation}
Y^{-1} - 1 = g^2 \frac{e^{-\alpha}}{16 \alpha}.
\end{equation}
Representative results of $Y$ at different values of $\mu / T$ are
shown in Fig.~\ref{fig:eff_mass_absolute}, which should
be compared with the analogous Fig.~\ref{fig:z_factor_absolute}
for the wavefunction renormalization factor.

\section{SUMMARY AND CONCLUSIONS}
In this paper we have used an 
FRG approach as well as quantum Monte Carlo simulations to study
the dilute Bose gas with contact interaction in two dimensions.
From the approximate analytical solution of the FRG flow equations we have been able
to obtain explicit analytic results 
for thermodynamic state functions as well as for quasi-particle properties 
in the vicinity of the QCP. 
Our results for the thermodynamics and the wavefunction renormalization 
are expected to be valid for general $\mu / T$ in the normal phase save for the 
classical critical region around the BKT transition, thus extending previous analytic 
results \cite{Sac94,Ran12} for the thermodynamic observables 
which considered only the special case
 $\mu = 0$. 
A comparison with experimental data \cite{Hun11,Yef11,Zha12} as well as with 
an alternative FRG approach based on the numerical solution of flow equations
obtained within the gradient expansion \cite{Ran12} 
shows good agreement with our expressions for the state functions 
when the dimensionless effective interaction is sufficiently small ($g \lesssim 0.1$), 
while the density even agrees up to $g \approx 1$.

To investigate the validity and the limitations of our FRG approach, we have also studied the spin-$1/2$ quantum $XY$ model close to the dilute Bose gas QCP using QMC simulations. It turns out that with our FRG approach we can predict the behavior of both density and compressibility even at relatively high temperatures, using only the effective 
ultraviolet cutoff $\Lambda_0$ of the continuum model as a free parameter.
In particular we were able to describe the numerical data for negative chemical potential analytically 
which has not been done before. 
For both cases of $\mu = 0$ and $\mu < 0$ we could also fit our analytical leading order results to the QMC data; however, this requires a rather large 
value of the ultraviolet cutoff $\Lambda_0$ and a modified bare mass. 
From the finite-size scaling of the density we have determined the correlation length 
which is consistent with the analytic predictions within error bars. 
Moreover, the finite-size scaling also defines a region of validity for the 
continuum description of the lattice model. Outside this region the exponential 
behavior in Eq.~(\ref{expfit}) breaks down and the correlation length is of order unity.
Interestingly, in that regime the numerical data always shows a minimum 
in the density over temperature and in the compressibility
as a function of $T$ for a given $\alpha$, which is not captured 
by the continuum description.

Finally, we have computed the wavefunction renormalization factor $Z$ 
for hard-core bosons numerically using the self-consistent $T$-matrix approximation which for $\mu = 0$ should be accurate  at high temperatures. 
Comparing the data to our analytic FRG result for $Z$ which is only valid at very small $T$, we find that an intermediate temperature range where both the truncated FRG and the self-consistent $T$-matrix approximation are accurate unfortunately does not exist.

\begin{acknowledgments}
This work was supported by the SFB Transregio 49 and the Transregio 173 
of the Deutsche Forschungsgemeinschaft (DFG) and the Allianz f\"{u}r 
Hochleistungsrechnen Rheinland-Pfalz (AHRP).
\end{acknowledgments}

\begin{appendix}

\appendix

\section{EXACT FRG FLOW EQUATIONS}
\label{sec:exact_frg_flow_equations}

\begin{figure}
\includegraphics[width=\linewidth]{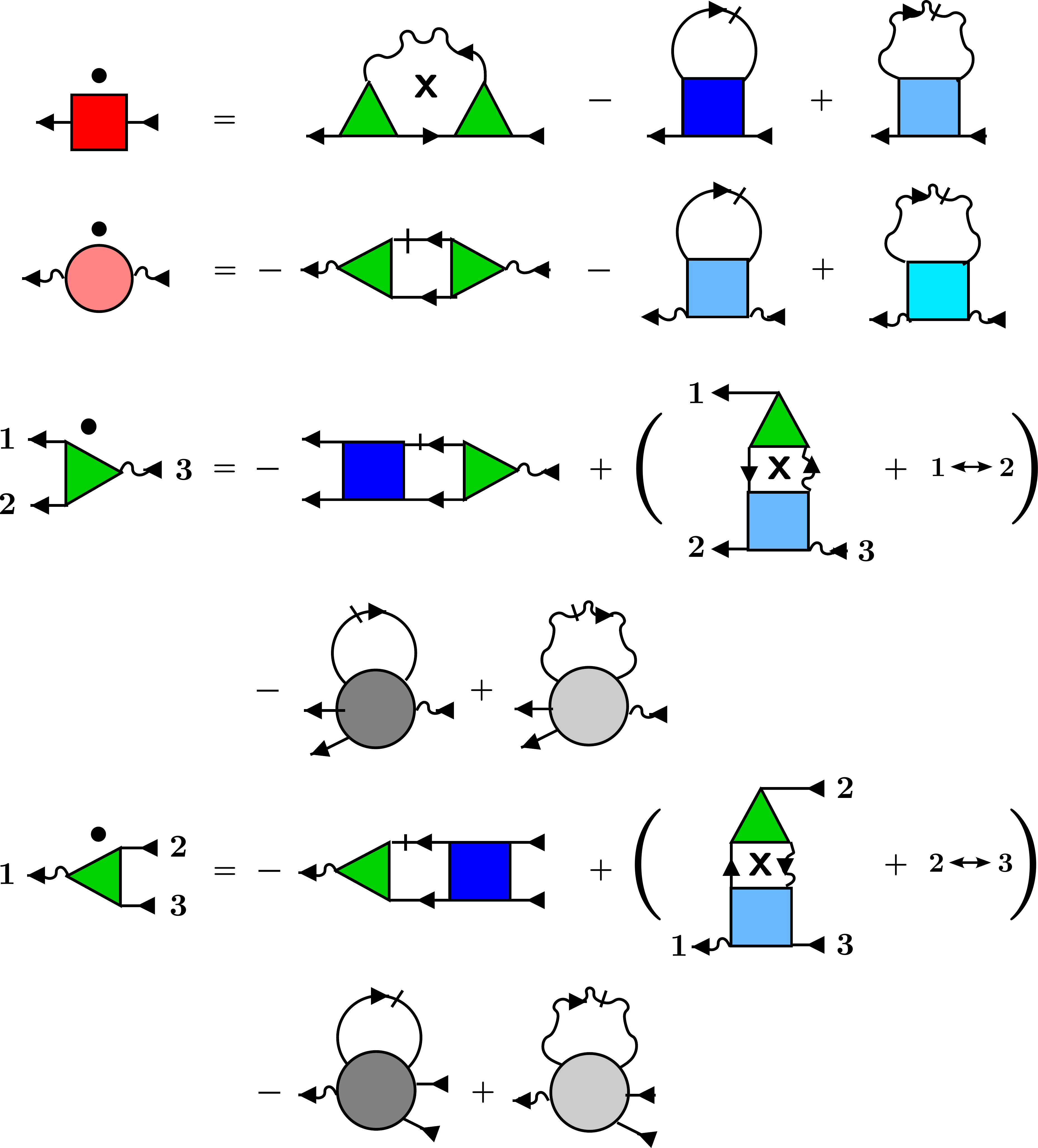}
\caption{
(Color online)
Graphical representation of the FRG flow equations (\ref{eq:flowself}-\ref{eq:gamma3b}) for the two- and three-point vertices, where the solid and wavy arrows denote the exact elementary and HS propagators, respectively. The interchange of labels applies to all diagrams inside the brackets, while a diagram containing a cross
stands for all different diagrams of this type with one of the internal propagators replaced by the corresponding single-scale propagator.
}
\label{fig:diagrams}
\end{figure}

In this appendix we write down  exact FRG flow equations for the one-line irreducible
vertices of the decoupled bosonic 
action in (\ref{eq:action_hs}). Therefore we modify the Gaussian propagators
of the elementary boson and the Hubbard-Stratonovich boson 
by inducing a cutoff $\Lambda$ which suppresses fluctuations with wave-vectors
smaller than $\Lambda$,
 \begin{align}
 G_{0} ( K ) &\rightarrow G_{0, \Lambda} ( K ),
 \\*
 F_{0} ( P ) &\rightarrow F_{0, \Lambda} ( P ).
 \end{align}
At some large initial value $\Lambda_0$ of the cutoff
the regularized bare action can be written in the following 
symmetrized form
 \begin{eqnarray}
 & & S_{\Lambda_0 } [ \bar{a} , a , \bar{\psi} , \psi ] 
 \\
& = & - \int_K G_{0 , \Lambda_0}^{-1} (K )
 \bar{a}_K  a_K 
 + \int_P  F_{0, \Lambda_0}^{-1} ( P ) \bar{\psi}_P \psi_P
% \nonumber
% \\
% &  & \hspace{-16mm} + \frac{1}{(2 ! )^2 } \int_{ K_1^{\prime}} \int_{ K_2^{\prime}}
% \int_{K_2 } \int_{ K_1} \delta_{ K_1^{\prime} + K_2^{\prime} , K_2 + K_1 }
% \nonumber
% \\
% & & \times
% \Gamma^{\bar{a} \bar{a} aa}_{\Lambda_0} ( K_1^{\prime} , K_2^{\prime} ;
% K_2 , K_1 )
% \bar{a}_{ K_1^{\prime}} \bar{a}_{ K_2^{\prime}} a_{K_2 } a_{ K_1 }
 \nonumber
 \\
 & +  &  \frac{1}{2 !} \int_{ K_1 }\int_{K_2}  \int_P
 \delta_{ K_1 + K_2 , P}  \bigl[
  \Gamma^{ \bar{a} \bar{a} {\psi}  }_{\Lambda_0} (  K_1 , K_2 ; P) 
  \bar{a}_{K_1} \bar{a}_{K_2 } {\psi}_P
 \nonumber
 \\
 & & \hspace{20mm} +
 \Gamma^{aa \bar{\psi} }_{\Lambda_0} (  K_1 , K_2 ; P) 
  a_{K_1} a_{K_2 }\bar{\psi}_P
 \bigr].
 \end{eqnarray}
Here the bare values of the symmetrized vertices are
 \begin{eqnarray}
%  & & \Gamma^{\bar{a} \bar{a} aa}_{\Lambda_0} ( K_1^{\prime} , K_2^{\prime} ;
% K_2 , K_1 ) = J_{\bd{k}_1^{\prime} - \bd{k}_1 } + J_{\bd{k}_1^{\prime} - \bd{k}_2 }
% \\
 & & 
 \Gamma^{  \bar{a} \bar{a}  {\psi}  }_{\Lambda_0} (  K_1 , K_2 ; P ) 
 =  \Gamma^{aa \bar{\psi}  }_{\Lambda_0} (  K_1 , K_2 ; P)  =    i .
 \end{eqnarray}
The exact FRG  equations, describing the flow of  one-line irreducible vertices
of the above theory as we reduce the cutoff, follow from the vertex expansion
of the FRG flow equation of the corresponding
generating functional.\cite{Wet93,Kop10}
The flowing inverse propagators are
of the form
 \begin{eqnarray}
 G^{-1}_{\Lambda} ( K ) & = & G_{0 , \Lambda}^{-1} (K ) - \Sigma_{\Lambda} ( K ),
 \\
 F^{-1}_{\Lambda} ( P ) & = & F_{0 , \Lambda}^{-1} (P ) + \Pi_{\Lambda} ( P ),
 \end{eqnarray}
where the self-energy $ \Sigma_{\Lambda} ( K )$ of the elementary boson 
satisfies the following exact flow equation,
 \begin{align}
 \partial_{\Lambda} \Sigma_{\Lambda} ( K ) &=
 \int_P \left[ \dot{F}_{\Lambda} ( P ) G_{\Lambda} ( P - K )
 +  {F}_{\Lambda} ( P ) \dot{G}_{\Lambda} ( P - K ) \right]
 \nonumber
 \\
& \hspace{6mm} \times  \Gamma^{ \bar{a} \bar{a} \psi  }_{\Lambda} ( P-K, K ; P )
 \Gamma^{ aa\bar{\psi} }_{\Lambda} (P-K,   K ; P )
 \nonumber
 \\
&- \int_{ K^{\prime}}
 \dot{G}_{\Lambda} ( K^{\prime} ) \Gamma^{\bar{a} \bar{a} aa}_{\Lambda}
 ( K , K^{\prime} ; K^{\prime} , K )
 \nonumber
 \\
 &+ \int_{ P}
 \dot{F}_{\Lambda} ( P ) \Gamma^{ \bar{a} a \bar{\psi} \psi}_{\Lambda}
 ( K; K; P; P ),
 \label{eq:flowself}
 \end{align}
which is shown graphically in the first line of Fig.~\ref{fig:diagrams}.
Here
$ \Gamma^{\bar{a} \bar{a} aa}_{\Lambda}
 ( K , K^{\prime} ; K^{\prime} , K )$
and  $\Gamma^{ \bar{a} a \bar{\psi} \psi}_{\Lambda}
 ( K; K; P; P )$
are one-line irreducible vertices with four external legs
of the type indicated by the superscripts, while
$ \dot{G}_{\Lambda} ( K )$ and $\dot{F}_{\Lambda} ( P )$
are the single-scale propagators \cite{Kop10} for the given
cutoff scheme; for example, for a sharp momentum  
cutoff the single-scale propagator
$ \dot{G}_{\Lambda} ( K )$ is
given in Eq.~(\ref{eq:Gdot}).

\begin{figure}
\includegraphics[width=\linewidth]{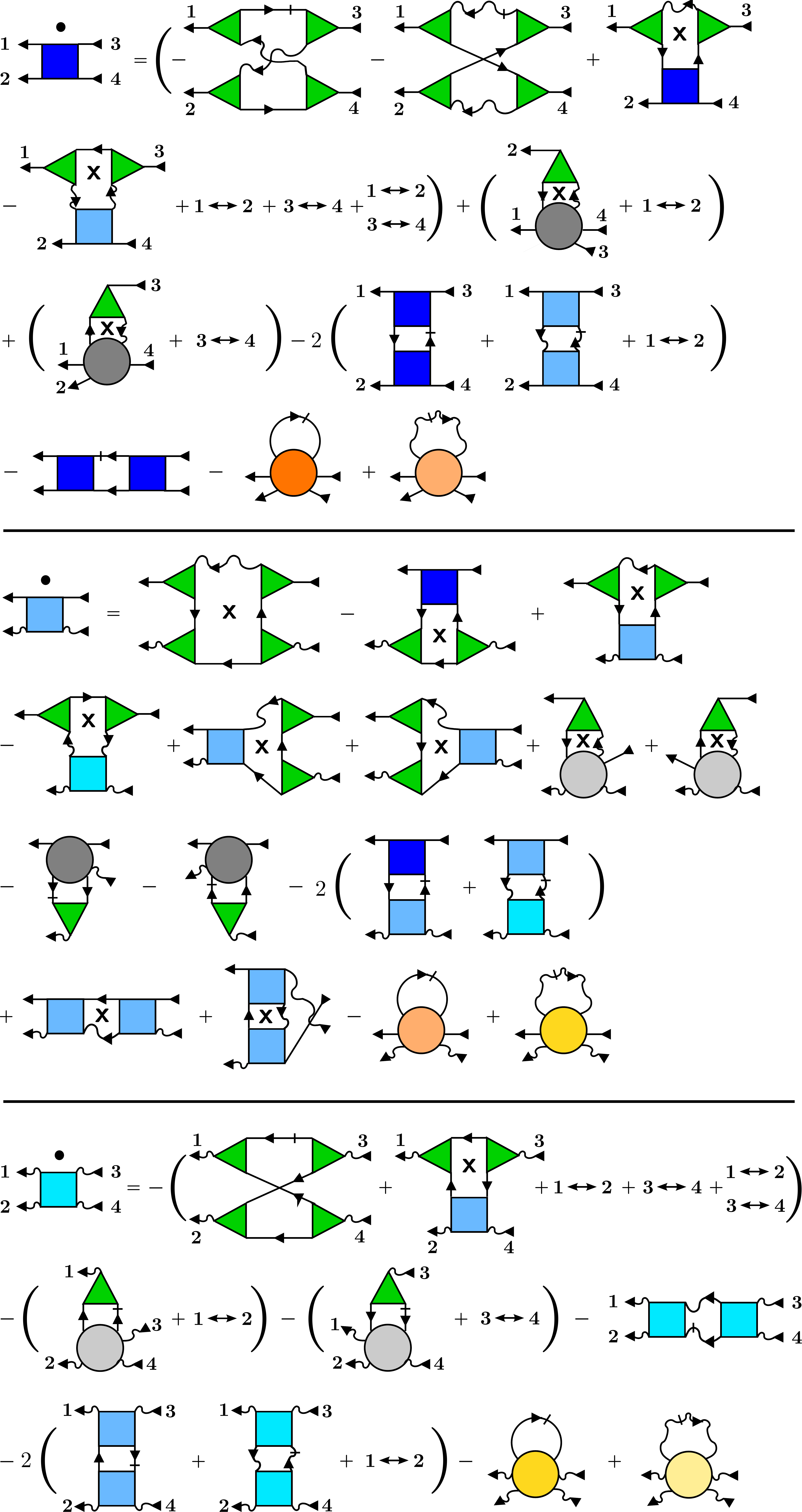}
\caption{
(Color online)
Graphical representation of the FRG flow equations for the four-point vertices $\Gamma^{\bar{a} \bar{a} a a}_{\Lambda}$, $\Gamma^{\bar{a} a \bar{\psi} \psi}_{\Lambda}$, and $\Gamma^{\bar{\psi} \bar{\psi} \psi \psi}_{\Lambda}$, where the solid and wavy arrows denote the exact elementary and HS propagators, respectively. The interchange of labels applies to all diagrams inside the brackets, while a diagram containing a  cross  represents 
all different diagrams of this type with one of the internal propagators replaced by the corresponding single-scale propagator.
}
\label{fig:diagrams_four}
\end{figure}

The  exact FRG flow equation for the self-energy
$ \Pi_{\Lambda} ( P )$ of the HS boson (which can be identified with the
irreducible particle-particle susceptibility) is
 \begin{align}
 \partial_{\Lambda} \Pi_{\Lambda} ( P ) &=
- \int_K  \dot{G}_{\Lambda} ( K ) G_{\Lambda} ( P - K )
 \Gamma^{ \bar{a} \bar{a}  \psi  }_{\Lambda} (P-K, K;  P )
\nonumber
 \\
 & \hspace{3mm} \times
 \Gamma^{aa \bar{\psi}  }_{\Lambda} ( P-K, K; P ) 
\nonumber
 \\
&- \int_{ K}
 \dot{G}_{\Lambda} ( K ) \Gamma^{ \bar{a} a    \bar{\psi} \psi   }_{\Lambda}
 ( K ; K ; P ; P )
 \nonumber
 \\
 &+ \int_{ P^{\prime}}
 \dot{F}_{\Lambda} ( P^{\prime} ) \Gamma^{\bar{\psi} \bar{\psi} \psi \psi}_{\Lambda}
 ( P , P^{\prime} ; P^{\prime} , P ).
 \label{eq:flowpol}
 \end{align}
This equation is shown graphically in the second line of
Fig.~\ref{fig:diagrams}.
The three-legged vertices satisfy the exact flow equations
 \begin{align}
 & \partial_{\Lambda} 
 \Gamma_{\Lambda}^{ \bar{a} \bar{a} \psi } (  K_1 , K_2 ; P) 
 \nonumber
 \\
 &=
- \int_K  \dot{G}_{\Lambda} ( K ) G_{\Lambda} ( P - K )
 \Gamma^{\bar{a} \bar{a} aa }_{\Lambda} ( K_1 , K_2 ; K, P-K )
\nonumber
\\
& \hspace{5mm} \times 
\Gamma^{ \bar{a} \bar{a} \psi  }_{\Lambda} ( P-K , K; P )
\nonumber
 \\
&+ \left[ \int_K  \dot{G}_{\Lambda} ( K ) F_{\Lambda} ( K_1 + K ) + G_{\Lambda} ( K ) \dot{F}_{\Lambda} ( K_1 + K ) \right]
\nonumber
\\
& \hspace{5mm} \times \Gamma^{\bar{a} a \bar{\psi} \psi }_{\Lambda} ( K_2; K; K_1 + K; P )
\Gamma^{ \bar{a} \bar{a} \psi  }_{\Lambda} ( K_1 , K; K_1 + K )
\nonumber
 \\
&+ \left[ \int_K  \dot{G}_{\Lambda} ( K ) F_{\Lambda} ( K_2 + K ) + G_{\Lambda} ( K ) \dot{F}_{\Lambda} ( K_2 + K ) \right]
\nonumber
\\
& \hspace{5mm} \times \Gamma^{\bar{a} a \bar{\psi} \psi }_{\Lambda} ( K_1; K; K_2 + K; P )
\Gamma^{ \bar{a} \bar{a} \psi  }_{\Lambda} ( K_2 , K; K_2 + K ) 
\nonumber
 \\ 
&- \int_{ K}
 \dot{G}_{\Lambda} ( K ) \Gamma^{ \bar{a} \bar{a}  \bar{a} a \psi }_{\Lambda}
 (  K_1 , K_2 , K ;  K ; P)
 \nonumber
 \\
 & + \int_{ P^{\prime}}
 \dot{F}_{\Lambda} ( P^{\prime} ) \Gamma^{ \bar{a} \bar{a}   \bar{\psi} \psi  \psi   }_{\Lambda}
 ( K_1, K_2; P^{\prime} ; P^{\prime}, P  )
 \label{eq:gamma3a}
 \end{align}
and
 \begin{align}
 & \partial_{\Lambda} 
 \Gamma_{\Lambda}^{ aa \bar{\psi}  } (  K_1 , K_2 ; P ) 
 \nonumber
 \\
 &=
- \int_K \dot{G}_{\Lambda} ( K ) G_{\Lambda} ( P - K )
 \Gamma^{\bar{a} \bar{a} aa }_{\Lambda} (  K, P-K ; K_1 , K_2)
\nonumber
 \\
 & \hspace{5mm} \times
 \Gamma^{aa \bar{\psi}  }_{\Lambda} ( P-K, K; P  )
\nonumber
 \\
&+ \left[ \int_K  \dot{G}_{\Lambda} ( K ) F_{\Lambda} ( K_1 + K ) + G_{\Lambda} ( K ) \dot{F}_{\Lambda} ( K_1 + K ) \right]
\nonumber
\\
& \hspace{5mm} \times \Gamma^{\bar{a} a \bar{\psi} \psi }_{\Lambda} ( K; K_2; P; K_1 + K )
\Gamma^{ a a \bar{\psi}  }_{\Lambda} ( K_1 , K; K_1 + K )
\nonumber
 \\
&+ \left[ \int_K  \dot{G}_{\Lambda} ( K ) F_{\Lambda} ( K_2 + K ) + G_{\Lambda} ( K ) \dot{F}_{\Lambda} ( K_2 + K ) \right]
\nonumber
\\
& \hspace{5mm} \times \Gamma^{\bar{a} a \bar{\psi} \psi }_{\Lambda} ( K; K_1; P; K_2 + K )
\Gamma^{ a a \bar{\psi}  }_{\Lambda} ( K_2 , K; K_2 + K )
\nonumber
 \\
&- \int_{ K}
 \dot{G}_{\Lambda} ( K ) \Gamma^{ {a} {a} a \bar{a} \bar{\psi}   }_{\Lambda}
 (  K_1 , K_2 , K ;  K ; P )
 \nonumber
 \\
 & + \int_{ P^{\prime}}
 \dot{F}_{\Lambda} ( P^{\prime} ) \Gamma^{ aa \psi \bar{\psi}   \bar{\psi}   }_{\Lambda}
 ( K_1 , K_2 ; P^{\prime} ;  P^{\prime}, P ).
 \label{eq:gamma3b}
 \end{align}
A graphical representation of these flow equations is shown in the lower half of Fig.~\ref{fig:diagrams}.
Because our action depends on two different types of fields corresponding to the
elementary boson and the HS boson, we have to keep track of three different types of four-point vertices.
Although in this work we do not need the exact flow equations of these vertices,
for later reference and for completeness we write down these flow equations in diagrammatic form in Fig.~\ref{fig:diagrams_four}.
Finally, let us also write down the exact FRG flow equation for the grand canonical 
potential $\Omega_\Lambda$,
 \begin{eqnarray}
 \frac{ \partial_{\Lambda} \Omega_{\Lambda}}{V} & = &
 - \int_K \frac{ \dot{G}_{0, \Lambda} ( K ) \Sigma_{\Lambda} ( K ) }{
 1 - G_{0, \Lambda} ( K ) \Sigma_{\Lambda} ( K ) }
 \nonumber
 \\
 & & +   \int_P \frac{ \dot{F}_{0, \Lambda} ( P ) \Pi_{\Lambda} ( P ) }{
 1 + F_{0, \Lambda} ( P ) \Pi_{\Lambda} ( P ) }.
 \end{eqnarray}
Note that with our normalization of the interaction, all combinatorial factors in the flow equations for the two- and three-point vertices are unity. Moreover, if we lump one minus sign into the HS propagators such that the combinations  $- F_{\Lambda} ( P )$ and $- \dot{F}_{\Lambda} ( P )$ appear everywhere, only a single overall minus sign multiplies all flow equations.

In the main text of this paper we introduce a sharp momentum cutoff scheme 
only in the propagator of the elementary boson.
In this scheme
\begin{align}
G_{0, \Lambda} (K) &= \frac{\Theta (|\bm{k}| - \Lambda)}{i \omega - \epsilon_{\bd{k}} + \mu},
\\
F_{0, \Lambda} (P) &= f_0,
\end{align}
so that the single-scale propagator for the HS boson vanishes identically in the exact flow equations given above,
\begin{equation}
\dot{F}_{\Lambda} (P) = 0.
\end{equation}

\section{ANALYTIC SOLUTION OF THE FLOW EQUATIONS}
 \label{sec:detailed_calcs}

\subsection{Self-energies}
\label{sec:detailed_calcs_self_energy}

In Sec.~\ref{sec:frg_qcp} we have obtained the following coupled RG flow equations
for the bosonic self-energies at vanishing external energies and momenta,
\begin{align}
\partial_l r_l &= \frac{1}{\frac{1}{u_0} + \tilde{\Pi}_l} \frac{4 \beta_l}{e^{\beta_l + r_l} - 1},
\label{eq:self_energy_coupled_again}
\\
\partial_l \tilde{\Pi}_l &= \frac{\beta_l}{\beta_l + r_l} \left[1 + \frac{2}{e^{\beta_l + r_l} - 1} \right],
\label{eq:pol_coupled_again}
\end{align}
where the boundary conditions have to be chosen such that the bosonic self-energies at the initial scale $\Lambda_0$ vanish, implying
\begin{equation}
r_0 = \alpha, \quad \tilde{\Pi}_0 = 0.
\end{equation}
It is clear from the flow equation of $r_l$ that it only starts to grow significantly when $\beta_l$ is of order unity, therefore the behavior of $\tilde{\Pi}_l$ before this point is not important. It turns out that
\begin{equation}
\partial_l \tilde{\Pi}_l = \frac{2 \beta_l e^{\beta_l + r}}{(e^{\beta_l + r} - 1)^2}
\label{eq:pi_approx_flow}
\end{equation}
with the boundary condition
\begin{equation}
\tilde{\Pi}_0 = \frac{1}{2} \ln \left[ \frac{\Lambda_0^2}{2 m (T - \mu)} \right]
\label{eq:boundary_condition_new}
\end{equation}
is a good replacement for the correct flow equation, leading to approximately the same flow for $r_l$; we will justify this approximation in the following.

First we note that the replacement $r_l \rightarrow r$ is good for any $r_0$: while for $r_0 \gtrsim 1$ the self-energy is negligible compared to $\mu / T$, we see from Eqs.~(\ref{eq:self_energy_coupled_again}) and (\ref{eq:pol_coupled_again}) that for $r_0 \ll 1$ the contribution of $r_l$ only becomes relevant when it is of the same order of magnitude as $\beta_l$ which is just when the flow of $r_l$ effectively stops. Secondly for large $\beta_l$ we can approximate the flow of $\tilde{\Pi}_l$ in Eq.~(\ref{eq:pol_coupled_again}) as
\begin{equation}
\partial_l \tilde{\Pi}_l = \frac{\beta_l}{\beta_l + r},
\label{eq:pi_approx_r}
\end{equation}
which is easily solved by
\begin{equation}
\tilde{\Pi}_l = \frac{1}{2} \ln \left( \frac{\beta_0 + r}{\beta_l + r} \right).
\end{equation}
Close to the QCP where $\beta_0$ is large we may extrapolate this result up to $\beta_l = 1$ as the corrections of order unity are assumed to be small compared to $\ln \beta_0$, yielding
\begin{equation}
\tilde{\Pi}_l\big|_{\beta_l = 1} \approx \frac{1}{2} \ln \left( \frac{\beta_0}{1 + \alpha} \right)
\end{equation}
for $\beta_0 \gg r$ and $\Sigma(0) / T \ll 1$, which agrees with the right-hand side of Eq.~(\ref{eq:boundary_condition_new}). Using Eq.~(\ref{eq:pi_approx_flow}) for $\tilde{\Pi}_l$ we find that it is almost constant for $\beta_l \gg 1$ (with irrelevant corrections at $\beta_l \gtrsim 1$), hence our replacement reproduces $\tilde{\Pi}_l$ at $\beta_l = 1$ quite well. Lastly we have to make sure that we also get the correct flow of $r_l$ for $\beta_l < 1$. According to Eq.~(\ref{eq:self_energy_coupled_again}) for $r_0 \gtrsim 1$ only the region $\beta_l \approx 1$ is relevant for the flow of $r_l$, thus we are left with the case $r_0 \ll 1$. Assuming $\beta_l \ll 1$ we may expand the exponentials in both versions of the flow equation for $\tilde{\Pi}_l$ which then coincide,
\begin{equation}
\partial_l \tilde{\Pi}_l = \frac{2 \beta_l}{(\beta_l + r)^2}.
\end{equation}
While there are deviations from the region $\beta_l \lesssim 1$, these are again small compared to $\ln \beta_0$. Therefore our approximate flow equation for $\tilde{\Pi}_l$ is justified for all relevant $r_0$ close to the QCP.

The advantage of Eq.~(\ref{eq:pi_approx_flow}) is that we can integrate this flow equation exactly,
\begin{equation}
\tilde{\Pi}_l = \frac{1}{e^{\beta_l + r} - 1} - \frac{1}{e^{\beta_0 + r} - 1} + \tilde{\Pi}_0 \approx \frac{1}{e^{\beta_l + r} - 1} + \tilde{\Pi}_0,
\end{equation}
so that the flow equation for $r_l$ reads
\begin{equation}
\partial_l r_l = \frac{2 g \beta_l}{e^{\beta_l + r} - 1 + \frac{g}{2}},
\label{eq:flowrsimplified_raw}
\end{equation}
where we have used $g$ from Eq.~(\ref{eq:g_def}). For $r \gtrsim 1$ we can readily drop the last term in the denominator of (\ref{eq:flowrsimplified_raw}), assuming that $g \ll 1$. If on the other hand $r \ll 1$ and $\beta_l \ll 1$ (which is then the relevant regime for the flow), we can expand the exponential in the denominator,
\begin{equation}
e^{\beta_l + r} - 1 + \frac{g}{2} \approx \beta_l + r + \frac{g}{2}.
\end{equation}
For $\mu \leq 0$ we will find that $r \gtrsim g W(1/g)$ for small $g$, hence we can again drop the last term in the denominator,
\begin{equation}
\partial_l r_l = \frac{2 g \beta_l}{e^{\beta_l + r} - 1}.
\label{eq:flowrsimplified}
\end{equation}
Integrating this and using $\beta_0 \gg 1$ we arrive at the transcendental equation
\begin{equation}
\frac{\Sigma (0)}{T} + g \ln \left( 1 - e^{- \frac{\Sigma (0)}{T} - \alpha} \right) = 0.
\end{equation}
After expanding $e^{\frac{\Sigma (0)}{T}} \approx 1 + \frac{\Sigma (0)}{T}$ we can solve for the self-energy and finally get
\begin{equation}
r = g W \left[ \frac{1}{g} \exp \left( \frac{e^\alpha - 1}{g} + \alpha \right) \right] - e^{\alpha} + 1 + \alpha
\label{eq:sigma_result_general_appendix}
\end{equation}
as given in Eq.~(\ref{eq:sigma_result_general}) in the main text.

We can also extend our calculation to positive $\mu$ by directly integrating Eq.~(\ref{eq:flowrsimplified_raw}), yielding
\begin{equation}
\frac{\Sigma (0)}{T} + \frac{2 g}{2 - 3 g} \ln \left[ e^{\frac{\Sigma (0)}{T}} - (1 - \frac{g}{2}) e^{-\alpha} \right] = 0,
\end{equation}
which we can again solve by expanding $e^{\frac{\Sigma (0)}{T}}$ to first order. The resulting expression,
\begin{align}
r &= \frac{2 g}{2 - 3 g} W \left[ \frac{2 - 3 g}{2 g} \exp\left( \frac{2 - 3 g}{2 g} \left[ 1 - e^{-\alpha} \left( 1 - \frac{g}{2} \right) \right] \right) \right]
\nonumber
\\
& \hspace{3mm} - 1 + e^{-\alpha} \left( 1 - \frac{g}{2} \right) + \alpha,
\label{eq:sigma_result_general_posMu_appendix}
\end{align}
is valid for all $\mu / T$ close to the QCP as long as we stay in the normal phase.

\subsection{Grand canonical potential}
\label{sec:detailed_free_energy}

To calculate the grand canonical potential $\Omega$ of the system within the FRG formalism we need to solve
the flow equation
\begin{equation}
\frac{\partial_\Lambda \Omega_\Lambda}{V} = - \int_K \frac{\dot{G}_{0,\Lambda} (K) \Sigma_\Lambda (K)}{1 - G_{0,\Lambda} (K) \Sigma_\Lambda (K)}.
\end{equation}
Within our sharp momentum cutoff scheme this reduces to
\begin{equation}
\frac{\partial_\Lambda \Omega_\Lambda}{V} = - \int_K \delta (k - \Lambda) \ln [1 - G_0 (K) \Sigma_\Lambda (K)].
\end{equation}
In the quasi-particle approximation we expand
\begin{equation}
\Sigma_\Lambda (K) \approx \Sigma_\Lambda (0) - (1 - Y_\Lambda^{-1}) \epsilon_{\bd{k}} + (1 - Z_\Lambda^{-1}) i \omega.
\end{equation}
We can then perform the momentum integration explicitly which only leaves us with the Matsubara sum. Rewriting it as a contour integral in the complex plane we find
\begin{align}
&\frac{\partial_\Lambda \Omega_\Lambda}{V} = \frac{T \Lambda}{2 \pi} \left(- \frac{1}{2 \pi i} \right) \int_\mathcal{C} \frac{dz}{e^z - 1}
\nonumber
\\*
&\times \ln \left[ 1 - \frac{\frac{\Sigma_\Lambda (0)}{T} - (1 - Y_\Lambda^{-1}) \beta_l + (1 - Z_\Lambda^{-1}) z}{z - \beta_l - \alpha} \right],
\end{align}
where we integrate in clockwise direction along two closed great half circles in the left and right complex half plane, respectively, which together encompass the whole complex plane without the imaginary axis. Here we have to require $\alpha > 0$ as will become clear in a moment. Then the integration over the left half plane vanishes as the function is holomorphic in this domain while the right contour encloses a branch cut. This may be seen by defining
\begin{equation}
z_1 = \beta_l + \alpha, \quad z_2 = Z_\Lambda (Y_\Lambda^{-1} \beta_l + r_l),
\end{equation}
and rewriting the argument of the above logarithm,
\begin{equation}
\frac{\partial_\Lambda \Omega_\Lambda}{V} = \frac{T \Lambda}{2 \pi} \left(- \frac{1}{2 \pi i} \right) \int_\mathcal{C} \frac{dz}{e^z - 1} \ln \left[ Z_\Lambda^{-1} \frac{z - z_2}{z - z_1} \right].
\end{equation}
Using the principal branch of the logarithm we find that it has a branch cut along the real axis, connecting the two points $z_1$ and $z_2$ where the logarithm diverges; due to the requirement $\alpha > 0$ we have $z_1, z_2 >0$. We can perform the integral by integrating alongside the branch cut,
\begin{align}
&\int_{z_2}^{z_1} \frac{dz}{e^z - 1} \ln \left[ Z_\Lambda^{-1} \frac{z + i \epsilon - z_2}{z + i \epsilon - z_1} \right] +
\nonumber
\\*
&\int_{z_1}^{z_2} \frac{dz}{e^z - 1} \ln \left[ Z_\Lambda^{-1} \frac{z - i \epsilon - z_2}{z - i \epsilon - z_1} \right] = - 2 \pi i \int_{z_2}^{z_1} \frac{dz}{e^z - 1},
\end{align}
where we have assumed that $z_1 > z_2$. However, if we 
repeat the calculation for the opposite case we obtain the same result. 
Switching to the logarithmic flow parameter $l$ and expressing the 
flow equation for $\Omega$ in terms of 
the reduced pressure $\tilde{p} = - \frac{\lambda_{\text{th}}^2}{T} \frac{\Omega}{V}$ we arrive at
\begin{equation}
\partial_l \tilde{p}_l = - 2 \beta_l \left[ \ln \left( \frac{e^{z_2} - 1}{e^{z_1} - 1} \right) - z_2 + z_1 \right].
\label{eq:flowptilde}
\end{equation}
At this point is it convenient to
introduce the flow parameter $b = \beta_l$ and set $Y_\Lambda = Z_\Lambda = 1$ for simplicity.
Then we obtain
\begin{align}
\partial_b \tilde{p}_b &= \ln \left( e^{b + r_b} - 1 \right) - b - r_b
\nonumber
\\
&- \left[ \ln \left( e^{b + \alpha} - 1 \right) - b - \alpha \right].
\end{align}
If we integrate the second line of this equation
and neglect terms of the order of $e^{- \beta_0}$ 
we obtain the initial condition of the reduced pressure,
\begin{equation}
- \int_0^\infty db \left[ \ln \left( e^{b + \alpha} - 1 \right) - b - \alpha \right] = \text{Li}_2 \left( e^{-\alpha} \right) = \tilde{p}_{\beta_0}.
\end{equation}
Hence in the physical limit $l \rightarrow \infty$  the reduced pressure is given by
\begin{equation}
\tilde{p} = - \int_0^\infty db \left[ \ln \left( e^{b + r_b} - 1 \right) - b - r_b \right].
\label{eq:integralEquationRedPressure}
\end{equation}
In the derivation of this equation we have assumed $\alpha > 0$ which ensures that the initial condition of the flow, the reduced pressure of the non-interacting system, exists. However, since (\ref{eq:integralEquationRedPressure}) is well defined for any $\alpha$ as long as we stay in the normal phase above the BKT transition, we can extend our result for $\tilde{p}$ to this region. In the simplest approximation where we replace $r_b$ by its final value $r$ we get
\begin{equation}
\tilde{p} = \text{Li}_2 \left( e^{-r} \right),
\label{eq:appendixSimpleResultP}
\end{equation}
which coincides with (\ref{eq:result_pressure_free}) in the main text.

%Note that using an additive Litim cutoff \cite{Lit01},
%%
%\begin{align}
%G_{0,\Lambda}^{-1} (K) &= G_0^{-1} (K) - R_\Lambda (\bd{k}),
%\nonumber
%\\*
%R_\Lambda (\bd{k}) &= \left( \frac{\Lambda^2}{2m} - \frac{k^2}{2m} \right) \Theta (\Lambda - %k),
%\end{align}
%%
%results in a simpler flow for $\tilde{p}_l$,
%%
%\begin{equation}
%\partial_l \tilde{p}_l = - 2 \beta_l^2 \left[ \frac{1}{e^{\beta_l + \alpha} - 1} - \frac{1}{e^{\beta_l + r_l} - 1} \right],
%\end{equation}
%%
%which is also well-defined for positive chemical potential in the normal phase. Hence, our previous requirement of $\alpha > 0$ was indeed related to the hard cutoff.

\subsection{Reduced pressure at $\mu = 0$ for $Y_\Lambda = Z_\Lambda = 1$}
\label{sec:detailed_free_energy_mu_0}

We can refine the result in (\ref{eq:appendixSimpleResultP}) by keeping track of the flow of $r_b$. 
As only the region $b < 1$ is relevant, we may expand the exponentials both in the flow 
equation (\ref{eq:flowptilde})
of $\tilde{p}_b$ and in the flow equation (\ref{eq:flowrsimplified}) of $r_b$,
\begin{align}
\partial_b \tilde{p}_b  & = \ln \left[ \frac{b + r_b}{b} \right] - r_b,
\\
\partial_b  r_b & = - \frac{g}{b + r}.
\end{align}
For $b > r_b$ we may further expand the logarithm,
\begin{equation}
\partial_b \tilde{p}_b  = \frac{r_b}{b} - r_b,
\end{equation}
while for $b < r_b$ the replacement $r_b \rightarrow r$ is valid. Integrating both regions we find
\begin{align}
\tilde{p} &= \frac{\pi^2}{6} - r \ln 4 + r^2 - g \Biggl\{ \frac{\pi^2}{12} - 1 + r - r \ln 4
\nonumber
\\*
&+ \ln \left( 1 + \frac{1}{r} \right) \left( 2 r - \ln r \right) + \text{Li}_2 \left( - \frac{1}{r} \right) \Biggr\}.
\label{eq:pspress_better_result_long}
\end{align}
Expanding this result in terms of $g$ where $r$ is taken from Eq.~(\ref{eq:sigma_mu_0}) we arrive at
\begin{equation}
\tilde{p} \sim \frac{\pi^2}{6} - \frac{g}{2} W^2(1/g).
\label{eq:pspressBetterShortAppen}
\end{equation}
Note that we can also obtain this result in a different way as follows: a perturbative expansion of the reduced pressure in the bare interaction $u_0$ yields to first order
\begin{equation}
\tilde{p} - \tilde{p}_0 \approx - u_0 \tilde{n}_0^2,
\end{equation}
where $\tilde{n}_0$ is the phase-space density of the non-interacting Bose gas. Evaluating the same two-loop diagram using dressed propagators and neglecting the momentum and frequency dependence of $\Sigma (K)$ and $\Pi (P)$ we obtain
\begin{equation}
\tilde{p} - \tilde{p}_0 \approx - u \tilde{n}^2.
\end{equation}
Inserting our FRG results for $\mu = 0$, where to leading order
\begin{equation}
u \equiv \lim_{l \to \infty} \frac{1}{\frac{1}{u_0} + \tilde{\Pi}_l} \sim \frac{g}{2}
\end{equation}
and
\begin{equation}
\tilde{n} \sim W(1/g),
\end{equation}
we again arrive at Eq.~(\ref{eq:pspressBetterShortAppen}).

\subsection{Wavefunction renormalization}
\label{sec:detailed_z_factor}

As the flow of $Z_\Lambda$ is given by
\begin{equation}
\partial_\Lambda Z_\Lambda^{-1} = - \partial_\Lambda \partial_{\omega} \Sigma_\Lambda (0, \omega + i 0^+) \bigr|_{\omega = 0},
\end{equation}
we need the frequency dependence of the self-energy and thus in turn $\tilde{\Pi}_\Lambda (0, i\bar{\omega})$. If we neglect the Bose function in its flow equation which is allowed for non-positive $\mu$ and replace $r_l \rightarrow r$ we get
\begin{equation}
\tilde{\Pi}_\Lambda (0, i\bar{\omega}) = \frac{1}{2} \ln \left[ \frac{2 \beta_0 + 2 r - \beta i \bar{\omega}}{2 \beta_l + 2 r - \beta i \bar{\omega}} \right].
\label{eq:pi_tilde_frequency_dependence}
\end{equation}
Evaluating the Matsubara sum in $\partial_\Lambda \Sigma_\Lambda$ we find that the additional pole due to (\ref{eq:pi_tilde_frequency_dependence}) is exponentially suppressed, so that we obtain
\begin{eqnarray}
& & \partial_l Z_l^{-1} = \frac{4 Z_l \beta_l}{e^{\beta_l + Z_l r_l} - 1} 
 \nonumber
 \\
 & \times &
\Bigg\{
\frac{1}{u_0^{-1} + \frac{1}{2} \ln \left( \frac{2\beta_0 - \beta_l + r}{\beta_l + r} \right)}  \frac{e^{\beta_l + Z_l r_l}}{e^{\beta_l + Z_l r_l} - 1}
\nonumber
\\*
& & + \Biggl[ \frac{1}{ u_0^{-1} + \frac{1}{2} \ln \left( \frac{2\beta_0 - \beta_l + r}{\beta_l + r} \right)}
 \Biggr]^2  \frac{\beta_0 - \beta_l}{(2\beta_0 - \beta_l + r) (\beta_l + r)} \Bigg\}.
\label{eq:z_factor_full_flow}
 \nonumber
 \\
 & &
\end{eqnarray}
We can simplify this by replacing $r_l \rightarrow r$, setting $Z_l = 1$ on the right hand side and realizing that the second term in the curly brackets is suppressed by a factor of $g$, which results in
\begin{align}
&\partial_l Z_l^{-1} = \frac{4 \beta_l e^{\beta_l + r}}{\left(e^{\beta_l + r} - 1\right)^2} \frac{1}{u_0^{-1} + \frac{1}{2} \ln \left( \frac{2\beta_0 - \beta_l + r}{\beta_l + r} \right)}.
\end{align}
Furthermore we may replace the second fraction in this expression by $g/2$ so that we arrive at
\begin{equation}
\partial_l Z_l^{-1} = \frac{2 g \beta_l e^{\beta_l + r}}{ \left( e^{\beta_l + r} - 1 \right)^2},
\end{equation}
which should be good for $\mu \leq 0$ in the limit of small $g$.

\end{appendix}

\bibliographystyle{apsrev4-1}
\bibliography{dilute-bosons}

%merlin.mbs apsrev4-1.bst 2010-07-25 4.21a (PWD, AO, DPC) hacked
%Control: key (0)
%Control: author (72) initials jnrlst
%Control: editor formatted (1) identically to author
%Control: production of article title (-1) disabled
%Control: page (0) single
%Control: year (1) truncated
%Control: production of eprint (0) enabled
\begin{thebibliography}{38}%
\makeatletter
\providecommand \@ifxundefined [1]{%
 \@ifx{#1\undefined}
}%
\providecommand \@ifnum [1]{%
 \ifnum #1\expandafter \@firstoftwo
 \else \expandafter \@secondoftwo
 \fi
}%
\providecommand \@ifx [1]{%
 \ifx #1\expandafter \@firstoftwo
 \else \expandafter \@secondoftwo
 \fi
}%
\providecommand \natexlab [1]{#1}%
\providecommand \enquote  [1]{``#1''}%
\providecommand \bibnamefont  [1]{#1}%
\providecommand \bibfnamefont [1]{#1}%
\providecommand \citenamefont [1]{#1}%
\providecommand \href@noop [0]{\@secondoftwo}%
\providecommand \href [0]{\begingroup \@sanitize@url \@href}%
\providecommand \@href[1]{\@@startlink{#1}\@@href}%
\providecommand \@@href[1]{\endgroup#1\@@endlink}%
\providecommand \@sanitize@url [0]{\catcode `\\12\catcode `\$12\catcode
  `\&12\catcode `\#12\catcode `\^12\catcode `\_12\catcode `\%12\relax}%
\providecommand \@@startlink[1]{}%
\providecommand \@@endlink[0]{}%
\providecommand \url  [0]{\begingroup\@sanitize@url \@url }%
\providecommand \@url [1]{\endgroup\@href {#1}{\urlprefix }}%
\providecommand \urlprefix  [0]{URL }%
\providecommand \Eprint [0]{\href }%
\providecommand \doibase [0]{http://dx.doi.org/}%
\providecommand \selectlanguage [0]{\@gobble}%
\providecommand \bibinfo  [0]{\@secondoftwo}%
\providecommand \bibfield  [0]{\@secondoftwo}%
\providecommand \translation [1]{[#1]}%
\providecommand \BibitemOpen [0]{}%
\providecommand \bibitemStop [0]{}%
\providecommand \bibitemNoStop [0]{.\EOS\space}%
\providecommand \EOS [0]{\spacefactor3000\relax}%
\providecommand \BibitemShut  [1]{\csname bibitem#1\endcsname}%
\let\auto@bib@innerbib\@empty
%</preamble>
\bibitem [{\citenamefont {Sachdev}(2011)}]{Sac11}%
  \BibitemOpen
  \bibfield  {author} {\bibinfo {author} {\bibfnamefont {S.}~\bibnamefont
  {Sachdev}},\ }\href@noop {} {\emph {\bibinfo {title} {Quantum Phase
  Transitions}}},\ \bibinfo {edition} {2nd}\ ed.\ (\bibinfo  {publisher}
  {Cambridge University Press, New York},\ \bibinfo {year} {2011})\BibitemShut
  {NoStop}%
\bibitem [{\citenamefont {Popov}(1972)}]{Pop72}%
  \BibitemOpen
  \bibfield  {author} {\bibinfo {author} {\bibfnamefont {V.~N.}\ \bibnamefont
  {Popov}},\ }\href {\doibase 10.1007/BF01028373} {\bibfield  {journal}
  {\bibinfo  {journal} {Theoret. and Math. Phys.}\ }\textbf {\bibinfo {volume}
  {11}},\ \bibinfo {pages} {565} (\bibinfo {year} {1972})}\BibitemShut
  {NoStop}%
\bibitem [{\citenamefont {Mermin}\ and\ \citenamefont {Wagner}(1966)}]{Mer66}%
  \BibitemOpen
  \bibfield  {author} {\bibinfo {author} {\bibfnamefont {N.~D.}\ \bibnamefont
  {Mermin}}\ and\ \bibinfo {author} {\bibfnamefont {H.}~\bibnamefont
  {Wagner}},\ }\href {\doibase 10.1103/PhysRevLett.17.1133} {\bibfield
  {journal} {\bibinfo  {journal} {Phys. Rev. Lett.}\ }\textbf {\bibinfo
  {volume} {17}},\ \bibinfo {pages} {1133} (\bibinfo {year}
  {1966})}\BibitemShut {NoStop}%
\bibitem [{\citenamefont {Berezinskii}(1971)}]{Ber71}%
  \BibitemOpen
  \bibfield  {author} {\bibinfo {author} {\bibfnamefont {V.~L.}\ \bibnamefont
  {Berezinskii}},\ }\href@noop {} {\bibfield  {journal} {\bibinfo  {journal}
  {Sov. Phys. JETP}\ }\textbf {\bibinfo {volume} {32}},\ \bibinfo {pages} {493}
  (\bibinfo {year} {1971})}\BibitemShut {NoStop}%
\bibitem [{\citenamefont {Berezinskii}(1972)}]{Ber72}%
  \BibitemOpen
  \bibfield  {author} {\bibinfo {author} {\bibfnamefont {V.~L.}\ \bibnamefont
  {Berezinskii}},\ }\href@noop {} {\bibfield  {journal} {\bibinfo  {journal}
  {Sov. Phys. JETP}\ }\textbf {\bibinfo {volume} {34}},\ \bibinfo {pages} {610}
  (\bibinfo {year} {1972})}\BibitemShut {NoStop}%
\bibitem [{\citenamefont {Kosterlitz}\ and\ \citenamefont
  {Thouless}(1973)}]{Kos73}%
  \BibitemOpen
  \bibfield  {author} {\bibinfo {author} {\bibfnamefont {J.~M.}\ \bibnamefont
  {Kosterlitz}}\ and\ \bibinfo {author} {\bibfnamefont {D.~J.}\ \bibnamefont
  {Thouless}},\ }\href@noop {} {\bibfield  {journal} {\bibinfo  {journal} {J.
  Phys. C}\ }\textbf {\bibinfo {volume} {6}},\ \bibinfo {pages} {1181}
  (\bibinfo {year} {1973})}\BibitemShut {NoStop}%
\bibitem [{\citenamefont {Kosterlitz}(1974)}]{Kos74}%
  \BibitemOpen
  \bibfield  {author} {\bibinfo {author} {\bibfnamefont {J.~M.}\ \bibnamefont
  {Kosterlitz}},\ }\href@noop {} {\bibfield  {journal} {\bibinfo  {journal} {J.
  Phys. C}\ }\textbf {\bibinfo {volume} {7}},\ \bibinfo {pages} {1046}
  (\bibinfo {year} {1974})}\BibitemShut {NoStop}%
\bibitem [{\citenamefont {Fisher}\ \emph {et~al.}(1989)\citenamefont {Fisher},
  \citenamefont {Weichman}, \citenamefont {Grinstein},\ and\ \citenamefont
  {Fisher}}]{Fis89}%
  \BibitemOpen
  \bibfield  {author} {\bibinfo {author} {\bibfnamefont {M.~P.~A.}\
  \bibnamefont {Fisher}}, \bibinfo {author} {\bibfnamefont {P.~B.}\
  \bibnamefont {Weichman}}, \bibinfo {author} {\bibfnamefont {G.}~\bibnamefont
  {Grinstein}}, \ and\ \bibinfo {author} {\bibfnamefont {D.~S.}\ \bibnamefont
  {Fisher}},\ }\href {\doibase 10.1103/PhysRevB.40.546} {\bibfield  {journal}
  {\bibinfo  {journal} {Phys. Rev. B}\ }\textbf {\bibinfo {volume} {40}},\
  \bibinfo {pages} {546} (\bibinfo {year} {1989})}\BibitemShut {NoStop}%
\bibitem [{\citenamefont {Sachdev}\ and\ \citenamefont {Dunkel}(2006)}]{Sac06}%
  \BibitemOpen
  \bibfield  {author} {\bibinfo {author} {\bibfnamefont {S.}~\bibnamefont
  {Sachdev}}\ and\ \bibinfo {author} {\bibfnamefont {E.~R.}\ \bibnamefont
  {Dunkel}},\ }\href {\doibase 10.1103/PhysRevB.73.085116} {\bibfield
  {journal} {\bibinfo  {journal} {Phys. Rev. B}\ }\textbf {\bibinfo {volume}
  {73}},\ \bibinfo {pages} {085116} (\bibinfo {year} {2006})}\BibitemShut
  {NoStop}%
\bibitem [{\citenamefont {Hadzibabic}\ \emph {et~al.}(2006)\citenamefont
  {Hadzibabic}, \citenamefont {Kr\"uger}, \citenamefont {Cheneau},
  \citenamefont {Battelier},\ and\ \citenamefont {Dalibard}}]{Had06}%
  \BibitemOpen
  \bibfield  {author} {\bibinfo {author} {\bibfnamefont {Z.}~\bibnamefont
  {Hadzibabic}}, \bibinfo {author} {\bibfnamefont {P.}~\bibnamefont
  {Kr\"uger}}, \bibinfo {author} {\bibfnamefont {M.}~\bibnamefont {Cheneau}},
  \bibinfo {author} {\bibfnamefont {B.}~\bibnamefont {Battelier}}, \ and\
  \bibinfo {author} {\bibfnamefont {J.}~\bibnamefont {Dalibard}},\ }\href
  {\doibase 10.1038/nature04851} {\bibfield  {journal} {\bibinfo  {journal}
  {Nature}\ }\textbf {\bibinfo {volume} {441}},\ \bibinfo {pages} {1118}
  (\bibinfo {year} {2006})}\BibitemShut {NoStop}%
\bibitem [{\citenamefont {Kr\"uger}\ \emph {et~al.}(2007)\citenamefont
  {Kr\"uger}, \citenamefont {Hadzibabic},\ and\ \citenamefont
  {Dalibard}}]{Kru07}%
  \BibitemOpen
  \bibfield  {author} {\bibinfo {author} {\bibfnamefont {P.}~\bibnamefont
  {Kr\"uger}}, \bibinfo {author} {\bibfnamefont {Z.}~\bibnamefont
  {Hadzibabic}}, \ and\ \bibinfo {author} {\bibfnamefont {J.}~\bibnamefont
  {Dalibard}},\ }\href {\doibase 10.1103/PhysRevLett.99.040402} {\bibfield
  {journal} {\bibinfo  {journal} {Phys. Rev. Lett.}\ }\textbf {\bibinfo
  {volume} {99}},\ \bibinfo {pages} {040402} (\bibinfo {year}
  {2007})}\BibitemShut {NoStop}%
\bibitem [{\citenamefont {Clad\'e}\ \emph {et~al.}(2009)\citenamefont
  {Clad\'e}, \citenamefont {Ryu}, \citenamefont {Ramanathan}, \citenamefont
  {Helmerson},\ and\ \citenamefont {Phillips}}]{Cla09}%
  \BibitemOpen
  \bibfield  {author} {\bibinfo {author} {\bibfnamefont {P.}~\bibnamefont
  {Clad\'e}}, \bibinfo {author} {\bibfnamefont {C.}~\bibnamefont {Ryu}},
  \bibinfo {author} {\bibfnamefont {A.}~\bibnamefont {Ramanathan}}, \bibinfo
  {author} {\bibfnamefont {K.}~\bibnamefont {Helmerson}}, \ and\ \bibinfo
  {author} {\bibfnamefont {W.~D.}\ \bibnamefont {Phillips}},\ }\href {\doibase
  10.1103/PhysRevLett.102.170401} {\bibfield  {journal} {\bibinfo  {journal}
  {Phys. Rev. Lett.}\ }\textbf {\bibinfo {volume} {102}},\ \bibinfo {pages}
  {170401} (\bibinfo {year} {2009})}\BibitemShut {NoStop}%
\bibitem [{\citenamefont {Tung}\ \emph {et~al.}(2010)\citenamefont {Tung},
  \citenamefont {Lamporesi}, \citenamefont {Lobser}, \citenamefont {Xia},\ and\
  \citenamefont {Cornell}}]{Tun10}%
  \BibitemOpen
  \bibfield  {author} {\bibinfo {author} {\bibfnamefont {S.}~\bibnamefont
  {Tung}}, \bibinfo {author} {\bibfnamefont {G.}~\bibnamefont {Lamporesi}},
  \bibinfo {author} {\bibfnamefont {D.}~\bibnamefont {Lobser}}, \bibinfo
  {author} {\bibfnamefont {L.}~\bibnamefont {Xia}}, \ and\ \bibinfo {author}
  {\bibfnamefont {E.~A.}\ \bibnamefont {Cornell}},\ }\href {\doibase
  10.1103/PhysRevLett.105.230408} {\bibfield  {journal} {\bibinfo  {journal}
  {Phys. Rev. Lett.}\ }\textbf {\bibinfo {volume} {105}},\ \bibinfo {pages}
  {230408} (\bibinfo {year} {2010})}\BibitemShut {NoStop}%
\bibitem [{\citenamefont {Hung}\ \emph {et~al.}(2011)\citenamefont {Hung},
  \citenamefont {Zhang}, \citenamefont {Gemelke},\ and\ \citenamefont
  {Chin}}]{Hun11}%
  \BibitemOpen
  \bibfield  {author} {\bibinfo {author} {\bibfnamefont {C.-L.}\ \bibnamefont
  {Hung}}, \bibinfo {author} {\bibfnamefont {X.}~\bibnamefont {Zhang}},
  \bibinfo {author} {\bibfnamefont {N.}~\bibnamefont {Gemelke}}, \ and\
  \bibinfo {author} {\bibfnamefont {C.}~\bibnamefont {Chin}},\ }\href {\doibase
  10.1038/nature09722} {\bibfield  {journal} {\bibinfo  {journal} {Nature}\
  }\textbf {\bibinfo {volume} {470}},\ \bibinfo {pages} {236} (\bibinfo {year}
  {2011})}\BibitemShut {NoStop}%
\bibitem [{\citenamefont {Yefsah}\ \emph {et~al.}(2011)\citenamefont {Yefsah},
  \citenamefont {Desbuquois}, \citenamefont {Chomaz}, \citenamefont
  {G\"unter},\ and\ \citenamefont {Dalibard}}]{Yef11}%
  \BibitemOpen
  \bibfield  {author} {\bibinfo {author} {\bibfnamefont {T.}~\bibnamefont
  {Yefsah}}, \bibinfo {author} {\bibfnamefont {R.}~\bibnamefont {Desbuquois}},
  \bibinfo {author} {\bibfnamefont {L.}~\bibnamefont {Chomaz}}, \bibinfo
  {author} {\bibfnamefont {K.~J.}\ \bibnamefont {G\"unter}}, \ and\ \bibinfo
  {author} {\bibfnamefont {J.}~\bibnamefont {Dalibard}},\ }\href {\doibase
  10.1103/PhysRevLett.107.130401} {\bibfield  {journal} {\bibinfo  {journal}
  {Phys. Rev. Lett.}\ }\textbf {\bibinfo {volume} {107}},\ \bibinfo {pages}
  {130401} (\bibinfo {year} {2011})}\BibitemShut {NoStop}%
\bibitem [{\citenamefont {Zhang}\ \emph {et~al.}(2012)\citenamefont {Zhang},
  \citenamefont {Hung}, \citenamefont {Tung},\ and\ \citenamefont
  {Chin}}]{Zha12}%
  \BibitemOpen
  \bibfield  {author} {\bibinfo {author} {\bibfnamefont {X.}~\bibnamefont
  {Zhang}}, \bibinfo {author} {\bibfnamefont {C.-L.}\ \bibnamefont {Hung}},
  \bibinfo {author} {\bibfnamefont {S.-K.}\ \bibnamefont {Tung}}, \ and\
  \bibinfo {author} {\bibfnamefont {C.}~\bibnamefont {Chin}},\ }\href {\doibase
  10.1126/science.1217990} {\bibfield  {journal} {\bibinfo  {journal}
  {Science}\ }\textbf {\bibinfo {volume} {335}},\ \bibinfo {pages} {1070}
  (\bibinfo {year} {2012})}\BibitemShut {NoStop}%
\bibitem [{\citenamefont {Ha}\ \emph {et~al.}(2013)\citenamefont {Ha},
  \citenamefont {Hung}, \citenamefont {Zhang}, \citenamefont {Eismann},
  \citenamefont {Tung},\ and\ \citenamefont {Chin}}]{Ha13}%
  \BibitemOpen
  \bibfield  {author} {\bibinfo {author} {\bibfnamefont {L.-C.}\ \bibnamefont
  {Ha}}, \bibinfo {author} {\bibfnamefont {C.-L.}\ \bibnamefont {Hung}},
  \bibinfo {author} {\bibfnamefont {X.}~\bibnamefont {Zhang}}, \bibinfo
  {author} {\bibfnamefont {U.}~\bibnamefont {Eismann}}, \bibinfo {author}
  {\bibfnamefont {S.-K.}\ \bibnamefont {Tung}}, \ and\ \bibinfo {author}
  {\bibfnamefont {C.}~\bibnamefont {Chin}},\ }\href {\doibase
  10.1103/PhysRevLett.110.145302} {\bibfield  {journal} {\bibinfo  {journal}
  {Phys. Rev. Lett.}\ }\textbf {\bibinfo {volume} {110}},\ \bibinfo {pages}
  {145302} (\bibinfo {year} {2013})}\BibitemShut {NoStop}%
\bibitem [{\citenamefont {Desbuquois}\ \emph {et~al.}(2014)\citenamefont
  {Desbuquois}, \citenamefont {Yefsah}, \citenamefont {Chomaz}, \citenamefont
  {Weitenberg}, \citenamefont {Corman}, \citenamefont {Nascimb\`ene},\ and\
  \citenamefont {Dalibard}}]{Des14}%
  \BibitemOpen
  \bibfield  {author} {\bibinfo {author} {\bibfnamefont {R.}~\bibnamefont
  {Desbuquois}}, \bibinfo {author} {\bibfnamefont {T.}~\bibnamefont {Yefsah}},
  \bibinfo {author} {\bibfnamefont {L.}~\bibnamefont {Chomaz}}, \bibinfo
  {author} {\bibfnamefont {C.}~\bibnamefont {Weitenberg}}, \bibinfo {author}
  {\bibfnamefont {L.}~\bibnamefont {Corman}}, \bibinfo {author} {\bibfnamefont
  {S.}~\bibnamefont {Nascimb\`ene}}, \ and\ \bibinfo {author} {\bibfnamefont
  {J.}~\bibnamefont {Dalibard}},\ }\href {\doibase
  10.1103/PhysRevLett.113.020404} {\bibfield  {journal} {\bibinfo  {journal}
  {Phys. Rev. Lett.}\ }\textbf {\bibinfo {volume} {113}},\ \bibinfo {pages}
  {020404} (\bibinfo {year} {2014})}\BibitemShut {NoStop}%
\bibitem [{\citenamefont {Fletcher}\ \emph {et~al.}(2015)\citenamefont
  {Fletcher}, \citenamefont {Robert{-}de{-}Saint{-}Vincent}, \citenamefont
  {Man}, \citenamefont {Navon}, \citenamefont {Smith}, \citenamefont
  {Viebahn},\ and\ \citenamefont {Hadzibabic}}]{Fle15}%
  \BibitemOpen
  \bibfield  {author} {\bibinfo {author} {\bibfnamefont {R.~J.}\ \bibnamefont
  {Fletcher}}, \bibinfo {author} {\bibfnamefont {M.}~\bibnamefont
  {Robert{-}de{-}Saint{-}Vincent}}, \bibinfo {author} {\bibfnamefont
  {J.}~\bibnamefont {Man}}, \bibinfo {author} {\bibfnamefont {N.}~\bibnamefont
  {Navon}}, \bibinfo {author} {\bibfnamefont {R.~P.}\ \bibnamefont {Smith}},
  \bibinfo {author} {\bibfnamefont {K.~G.~H.}\ \bibnamefont {Viebahn}}, \ and\
  \bibinfo {author} {\bibfnamefont {Z.}~\bibnamefont {Hadzibabic}},\ }\href
  {\doibase 10.1103/PhysRevLett.114.255302} {\bibfield  {journal} {\bibinfo
  {journal} {Phys. Rev. Lett.}\ }\textbf {\bibinfo {volume} {114}},\ \bibinfo
  {pages} {255302} (\bibinfo {year} {2015})}\BibitemShut {NoStop}%
\bibitem [{\citenamefont {Fisher}\ and\ \citenamefont
  {Hohenberg}(1988)}]{Fis88}%
  \BibitemOpen
  \bibfield  {author} {\bibinfo {author} {\bibfnamefont {D.~S.}\ \bibnamefont
  {Fisher}}\ and\ \bibinfo {author} {\bibfnamefont {P.~C.}\ \bibnamefont
  {Hohenberg}},\ }\href {\doibase 10.1103/PhysRevB.37.4936} {\bibfield
  {journal} {\bibinfo  {journal} {Phys. Rev. B}\ }\textbf {\bibinfo {volume}
  {37}},\ \bibinfo {pages} {4936} (\bibinfo {year} {1988})}\BibitemShut
  {NoStop}%
\bibitem [{\citenamefont {Sachdev}\ \emph {et~al.}(1994)\citenamefont
  {Sachdev}, \citenamefont {Senthil},\ and\ \citenamefont {Shankar}}]{Sac94}%
  \BibitemOpen
  \bibfield  {author} {\bibinfo {author} {\bibfnamefont {S.}~\bibnamefont
  {Sachdev}}, \bibinfo {author} {\bibfnamefont {T.}~\bibnamefont {Senthil}}, \
  and\ \bibinfo {author} {\bibfnamefont {R.}~\bibnamefont {Shankar}},\ }\href
  {\doibase 10.1103/PhysRevB.50.258} {\bibfield  {journal} {\bibinfo  {journal}
  {Phys. Rev. B}\ }\textbf {\bibinfo {volume} {50}},\ \bibinfo {pages} {258}
  (\bibinfo {year} {1994})}\BibitemShut {NoStop}%
\bibitem [{\citenamefont {Prokof'ev}\ \emph {et~al.}(2001)\citenamefont
  {Prokof'ev}, \citenamefont {Ruebenacker},\ and\ \citenamefont
  {Svistunov}}]{Pro01}%
  \BibitemOpen
  \bibfield  {author} {\bibinfo {author} {\bibfnamefont {N.}~\bibnamefont
  {Prokof'ev}}, \bibinfo {author} {\bibfnamefont {O.}~\bibnamefont
  {Ruebenacker}}, \ and\ \bibinfo {author} {\bibfnamefont {B.}~\bibnamefont
  {Svistunov}},\ }\href {\doibase 10.1103/PhysRevLett.87.270402} {\bibfield
  {journal} {\bibinfo  {journal} {Phys. Rev. Lett.}\ }\textbf {\bibinfo
  {volume} {87}},\ \bibinfo {pages} {270402} (\bibinfo {year}
  {2001})}\BibitemShut {NoStop}%
\bibitem [{\citenamefont {Prokof'ev}\ and\ \citenamefont
  {Svistunov}(2002)}]{Pro02}%
  \BibitemOpen
  \bibfield  {author} {\bibinfo {author} {\bibfnamefont {N.}~\bibnamefont
  {Prokof'ev}}\ and\ \bibinfo {author} {\bibfnamefont {B.}~\bibnamefont
  {Svistunov}},\ }\href {\doibase 10.1103/PhysRevA.66.043608} {\bibfield
  {journal} {\bibinfo  {journal} {Phys. Rev. A}\ }\textbf {\bibinfo {volume}
  {66}},\ \bibinfo {pages} {043608} (\bibinfo {year} {2002})}\BibitemShut
  {NoStop}%
\bibitem [{\citenamefont {Bernardet}\ \emph {et~al.}(2002)\citenamefont
  {Bernardet}, \citenamefont {Batrouni}, \citenamefont {Meunier}, \citenamefont
  {Schmid}, \citenamefont {Troyer},\ and\ \citenamefont {Dorneich}}]{Ber02}%
  \BibitemOpen
  \bibfield  {author} {\bibinfo {author} {\bibfnamefont {K.}~\bibnamefont
  {Bernardet}}, \bibinfo {author} {\bibfnamefont {G.~G.}\ \bibnamefont
  {Batrouni}}, \bibinfo {author} {\bibfnamefont {J.-L.}\ \bibnamefont
  {Meunier}}, \bibinfo {author} {\bibfnamefont {G.}~\bibnamefont {Schmid}},
  \bibinfo {author} {\bibfnamefont {M.}~\bibnamefont {Troyer}}, \ and\ \bibinfo
  {author} {\bibfnamefont {A.}~\bibnamefont {Dorneich}},\ }\href {\doibase
  10.1103/PhysRevB.65.104519} {\bibfield  {journal} {\bibinfo  {journal} {Phys.
  Rev. B}\ }\textbf {\bibinfo {volume} {65}},\ \bibinfo {pages} {104519}
  (\bibinfo {year} {2002})}\BibitemShut {NoStop}%
\bibitem [{\citenamefont {Pilati}\ \emph {et~al.}(2005)\citenamefont {Pilati},
  \citenamefont {Boronat}, \citenamefont {Casulleras},\ and\ \citenamefont
  {Giorgini}}]{Pil05}%
  \BibitemOpen
  \bibfield  {author} {\bibinfo {author} {\bibfnamefont {S.}~\bibnamefont
  {Pilati}}, \bibinfo {author} {\bibfnamefont {J.}~\bibnamefont {Boronat}},
  \bibinfo {author} {\bibfnamefont {J.}~\bibnamefont {Casulleras}}, \ and\
  \bibinfo {author} {\bibfnamefont {S.}~\bibnamefont {Giorgini}},\ }\href
  {\doibase 10.1103/PhysRevA.71.023605} {\bibfield  {journal} {\bibinfo
  {journal} {Phys. Rev. A}\ }\textbf {\bibinfo {volume} {71}},\ \bibinfo
  {pages} {023605} (\bibinfo {year} {2005})}\BibitemShut {NoStop}%
\bibitem [{\citenamefont {Ran{\c c}on}\ and\ \citenamefont
  {Dupuis}(2012)}]{Ran12}%
  \BibitemOpen
  \bibfield  {author} {\bibinfo {author} {\bibfnamefont {A.}~\bibnamefont
  {Ran{\c c}on}}\ and\ \bibinfo {author} {\bibfnamefont {N.}~\bibnamefont
  {Dupuis}},\ }\href {\doibase 10.1103/PhysRevA.85.063607} {\bibfield
  {journal} {\bibinfo  {journal} {Phys. Rev. A}\ }\textbf {\bibinfo {volume}
  {85}},\ \bibinfo {pages} {063607} (\bibinfo {year} {2012})}\BibitemShut
  {NoStop}%
\bibitem [{\citenamefont {Ceccarelli}\ \emph {et~al.}(2013)\citenamefont
  {Ceccarelli}, \citenamefont {Nespolo}, \citenamefont {Pelissetto},\ and\
  \citenamefont {Vicari}}]{Cec13}%
  \BibitemOpen
  \bibfield  {author} {\bibinfo {author} {\bibfnamefont {G.}~\bibnamefont
  {Ceccarelli}}, \bibinfo {author} {\bibfnamefont {J.}~\bibnamefont {Nespolo}},
  \bibinfo {author} {\bibfnamefont {A.}~\bibnamefont {Pelissetto}}, \ and\
  \bibinfo {author} {\bibfnamefont {E.}~\bibnamefont {Vicari}},\ }\href
  {\doibase 10.1103/PhysRevB.88.024517} {\bibfield  {journal} {\bibinfo
  {journal} {Phys. Rev. B}\ }\textbf {\bibinfo {volume} {88}},\ \bibinfo
  {pages} {024517} (\bibinfo {year} {2013})}\BibitemShut {NoStop}%
\bibitem [{\citenamefont {Strassel}\ \emph {et~al.}(2015)\citenamefont
  {Strassel}, \citenamefont {Kopietz},\ and\ \citenamefont {Eggert}}]{St15}%
  \BibitemOpen
  \bibfield  {author} {\bibinfo {author} {\bibfnamefont {D.}~\bibnamefont
  {Strassel}}, \bibinfo {author} {\bibfnamefont {P.}~\bibnamefont {Kopietz}}, \
  and\ \bibinfo {author} {\bibfnamefont {S.}~\bibnamefont {Eggert}},\ }\href
  {\doibase 10.1103/PhysRevB.91.134406} {\bibfield  {journal} {\bibinfo
  {journal} {Phys. Rev. B}\ }\textbf {\bibinfo {volume} {91}},\ \bibinfo
  {pages} {134406} (\bibinfo {year} {2015})}\BibitemShut {NoStop}%
\bibitem [{\citenamefont {Kopietz}\ \emph {et~al.}(2010)\citenamefont
  {Kopietz}, \citenamefont {Bartosch},\ and\ \citenamefont {Sch\"utz}}]{Kop10}%
  \BibitemOpen
  \bibfield  {author} {\bibinfo {author} {\bibfnamefont {P.}~\bibnamefont
  {Kopietz}}, \bibinfo {author} {\bibfnamefont {L.}~\bibnamefont {Bartosch}}, \
  and\ \bibinfo {author} {\bibfnamefont {F.}~\bibnamefont {Sch\"utz}},\
  }\href@noop {} {\emph {\bibinfo {title} {Introduction to the Functional
  Renormalization Group}}}\ (\bibinfo  {publisher} {Springer, Berlin},\
  \bibinfo {year} {2010})\BibitemShut {NoStop}%
\bibitem [{\citenamefont {Stoof}\ and\ \citenamefont {Bijlsma}(1993)}]{Sto93}%
  \BibitemOpen
  \bibfield  {author} {\bibinfo {author} {\bibfnamefont {H.~T.~C.}\
  \bibnamefont {Stoof}}\ and\ \bibinfo {author} {\bibfnamefont
  {M.}~\bibnamefont {Bijlsma}},\ }\href {\doibase 10.1103/PhysRevE.47.939}
  {\bibfield  {journal} {\bibinfo  {journal} {Phys. Rev. E}\ }\textbf {\bibinfo
  {volume} {47}},\ \bibinfo {pages} {939} (\bibinfo {year} {1993})}\BibitemShut
  {NoStop}%
\bibitem [{\citenamefont {Streib}\ and\ \citenamefont {Kopietz}(2015)}]{Str15}%
  \BibitemOpen
  \bibfield  {author} {\bibinfo {author} {\bibfnamefont {S.}~\bibnamefont
  {Streib}}\ and\ \bibinfo {author} {\bibfnamefont {P.}~\bibnamefont
  {Kopietz}},\ }\href {\doibase 10.1103/PhysRevB.92.094442} {\bibfield
  {journal} {\bibinfo  {journal} {Phys. Rev. B}\ }\textbf {\bibinfo {volume}
  {92}},\ \bibinfo {pages} {094442} (\bibinfo {year} {2015})}\BibitemShut
  {NoStop}%
\bibitem [{\citenamefont {Fauseweh}\ \emph {et~al.}(2014)\citenamefont
  {Fauseweh}, \citenamefont {Stolze},\ and\ \citenamefont {Uhrig}}]{Fau14}%
  \BibitemOpen
  \bibfield  {author} {\bibinfo {author} {\bibfnamefont {B.}~\bibnamefont
  {Fauseweh}}, \bibinfo {author} {\bibfnamefont {J.}~\bibnamefont {Stolze}}, \
  and\ \bibinfo {author} {\bibfnamefont {G.~S.}\ \bibnamefont {Uhrig}},\ }\href
  {\doibase 10.1103/PhysRevB.90.024428} {\bibfield  {journal} {\bibinfo
  {journal} {Phys. Rev. B}\ }\textbf {\bibinfo {volume} {90}},\ \bibinfo
  {pages} {024428} (\bibinfo {year} {2014})}\BibitemShut {NoStop}%
\bibitem [{\citenamefont {Corless}\ \emph {et~al.}(1996)\citenamefont
  {Corless}, \citenamefont {Gonnet}, \citenamefont {Hare}, \citenamefont
  {Jeffrey},\ and\ \citenamefont {Knuth}}]{Cor96}%
  \BibitemOpen
  \bibfield  {author} {\bibinfo {author} {\bibfnamefont {R.~M.}\ \bibnamefont
  {Corless}}, \bibinfo {author} {\bibfnamefont {G.~H.}\ \bibnamefont {Gonnet}},
  \bibinfo {author} {\bibfnamefont {D.~E.~G.}\ \bibnamefont {Hare}}, \bibinfo
  {author} {\bibfnamefont {D.~J.}\ \bibnamefont {Jeffrey}}, \ and\ \bibinfo
  {author} {\bibfnamefont {D.~E.}\ \bibnamefont {Knuth}},\ }\href {\doibase
  10.1007/BF02124750} {\bibfield  {journal} {\bibinfo  {journal} {Adv. Comput.
  Math.}\ }\textbf {\bibinfo {volume} {5}},\ \bibinfo {pages} {329} (\bibinfo
  {year} {1996})}\BibitemShut {NoStop}%
\bibitem [{\citenamefont {Popov}(1983)}]{Pop83}%
  \BibitemOpen
  \bibfield  {author} {\bibinfo {author} {\bibfnamefont {V.~N.}\ \bibnamefont
  {Popov}},\ }\href@noop {} {\emph {\bibinfo {title} {Functional integrals in
  quantum field theory and statistical physics}}}\ (\bibinfo  {publisher} {D.
  Reidel, Dordrecht},\ \bibinfo {year} {1983})\BibitemShut {NoStop}%
\bibitem [{\citenamefont {Sylju\aa{}sen}\ and\ \citenamefont
  {Sandvik}(2002)}]{Sandvik2002}%
  \BibitemOpen
  \bibfield  {author} {\bibinfo {author} {\bibfnamefont {O.~F.}\ \bibnamefont
  {Sylju\aa{}sen}}\ and\ \bibinfo {author} {\bibfnamefont {A.~W.}\ \bibnamefont
  {Sandvik}},\ }\href {\doibase 10.1103/PhysRevE.66.046701} {\bibfield
  {journal} {\bibinfo  {journal} {Phys. Rev. E}\ }\textbf {\bibinfo {volume}
  {66}},\ \bibinfo {pages} {046701} (\bibinfo {year} {2002})}\BibitemShut
  {NoStop}%
\bibitem [{\citenamefont {Matsumoto}\ and\ \citenamefont
  {Nishimura}(1998)}]{Mersenne1998}%
  \BibitemOpen
  \bibfield  {author} {\bibinfo {author} {\bibfnamefont {M.}~\bibnamefont
  {Matsumoto}}\ and\ \bibinfo {author} {\bibfnamefont {T.}~\bibnamefont
  {Nishimura}},\ }\href {\doibase 10.1145/272991.272995} {\bibfield  {journal}
  {\bibinfo  {journal} {ACM Trans. Model. Comput. Simul.}\ }\textbf {\bibinfo
  {volume} {8}},\ \bibinfo {pages} {3} (\bibinfo {year} {1998})}\BibitemShut
  {NoStop}%
\bibitem [{\citenamefont {Ran{\c c}on}(2014)}]{Ran14}%
  \BibitemOpen
  \bibfield  {author} {\bibinfo {author} {\bibfnamefont {A.}~\bibnamefont
  {Ran{\c c}on}},\ }\href {\doibase 10.1103/PhysRevB.89.214418} {\bibfield
  {journal} {\bibinfo  {journal} {Phys. Rev. B}\ }\textbf {\bibinfo {volume}
  {89}},\ \bibinfo {pages} {214418} (\bibinfo {year} {2014})}\BibitemShut
  {NoStop}%
\bibitem [{\citenamefont {Wetterich}(1993)}]{Wet93}%
  \BibitemOpen
  \bibfield  {author} {\bibinfo {author} {\bibfnamefont {C.}~\bibnamefont
  {Wetterich}},\ }\href {\doibase
  http://dx.doi.org/10.1016/0370-2693(93)90726-X} {\bibfield  {journal}
  {\bibinfo  {journal} {Phys. Lett. B}\ }\textbf {\bibinfo {volume} {301}},\
  \bibinfo {pages} {90 } (\bibinfo {year} {1993})}\BibitemShut {NoStop}%
\end{thebibliography}%

\end{document}